\documentclass[preprint,aps,nofootinbib,amsmath,amssymb]{revtex4}
\usepackage[dvipdfm]{graphicx}
\usepackage{dcolumn}
\newcommand{ \slashchar }[1]{\setbox0=\hbox{$#1$}   
   \dimen0=\wd0                                     
   \setbox1=\hbox{/} \dimen1=\wd1                   
   \ifdim\dimen0>\dimen1                            
      \rlap{\hbox to \dimen0{\hfil/\hfil}}          
      #1                                            
   \else                                            
      \rlap{\hbox to \dimen1{\hfil$#1$\hfil}}       
      /                                             
   \fi}                                             %
\usepackage{amsmath}
\renewcommand{\Re}{\mathop{\mbox{Re}}} 
\renewcommand{\Im}{\mathop{\mbox{Im}}} 
\newcommand{\calH}{{\cal H}}

\newcommand{\calO}{{\cal O}}

\newcommand{\bra}[1]{{\langle {#1} |}}
\newcommand{\ket}[1]{{| {#1} \rangle}}
\renewcommand{\l}{\ell}

\newcommand{\qbar}{\bar{q}}
\newcommand{\lbar}{\bar{\l}}
\newcommand{\ubar}{\bar{u}}
\newcommand{\dbar}{\bar{d}}

\newcommand{\sbar}{\bar{s}}

\newcommand{\barB}{\bar{B}}
\newcommand{\barK}{\bar{K}}

\newcommand{\MeV}{\text{ MeV}}
\newcommand{\GeV}{\text{ GeV}}

\newcommand{\ps}{\text{ ps}}
\newcommand{\degree}{{}^\circ}

\newcommand{\g}{\text{\sl g}}
\newcommand{\mhat}{\hat{m}}
\begin{document}
\title{Pseudoscalar and Scalar Operators of Higgs-Penguins in the MSSM
and $B\to \phi K^{*},\,K\eta^{(}{}'{}^{)}$ Decays}
\author{Hisaki Hatanaka} \author{Kwei-Chou Yang}
\affiliation{ Department of Physics, Chung-Yuan Christian University,
Chungli, Taiwan 320, R.O.C.}
\date{December 28, 2007}
\preprint{arXiv:0711.3086 [hep-ph]}
%
%
%
\begin{abstract}
We study the effect of $b\to s\sbar s$ scalar/pseudoscalar operators
in $B\to K\eta^{(\prime)},\,\phi K^*$ decays.
In the minimal supersymmetric standard model (MSSM), such
scalar/pseudoscalar operators can be induced by the penguin diagrams of
neutral Higgs bosons.
These operators can be Fierz-transformed into tensor operators, and
the resultant tensor operators could affect the transverse
polarization amplitudes in $B\to \phi K^*$ decays.
A combined analysis of the decays $B\to \phi K^*$ and $B \to K\eta^{(\prime)}$,
including $b\to s\bar{s}s$ scalar/pseudoscalar operators and their
Fierz-transformed tensor operators originated from the MSSM,
is performed.
Our study is based on the followings:
(1) Assuming that weak annihilations in $B\to\phi K^*$ is negligible and
the polarization puzzle is resolved by Fierz-transformed tensor
operators,
it results in too large coefficients of scalar/pseudoscalar operators, such
 that the resulting $B\to K\eta^{(\prime)}$ branching fractions are much
 larger than observations.
(2) When we take the weak annihilations in $B \to \phi K^*$ into
account, the polarization puzzle can be resolved.
In this case, new physics effects are strongly suppressed
and no more relevant to the enhancement of the transverse modes
in $B\to \phi K^*$ decays.
\end{abstract}
\maketitle
%
\newcommand{\NP}{{\rm NP}}
\newcommand{\Kstar}{{K^{*}}}
\newcommand{\Kstarz}{{K^{*0}}}
\newcommand{\Kstarm}{{K^{*-}}}
\newcommand{\Kstarp}{{K^{*+}}}
\newcommand{\Br}{{\cal B}}
\newcommand{\micro}{\times 10^6}
\newcommand{\ACP}{A_{CP}}
\newcommand{\SCP}[1]{{-\eta_{CP} S_{{#1}}}}
\newcommand{\phieta}{{\phi_\eta}}
\newcommand{\para}{\parallel}
\newcommand{\barA}{\bar{A}}
\newcommand{\barH}{\bar{H}}

\newcommand{\barKstar}{\bar{K}^{*}}
\newcommand{\barKstarz}{\bar{K}^{*0}}
\newcommand{\BABAR}{\text{BaBar} }
\newcommand{\BELLE}{\text{Belle} }
\newcommand{\PR}{(1+\gamma_5)}
\newcommand{\PL}{(1-\gamma_5)}
\newcommand{\etaP}{\eta^{(\prime)}}
\newcommand{\barKz}{\bar{K}^0}
\newcommand{\orderof}[1]{{{\calO}({#1})}}
\newcommand{\theratio}{(B-A)/B}
\newcommand{\chisq}{\chi^2}
\newcommand{\chimin}{{\chi^2}_{\min}}
\newcommand{\dof}{\text{d.o.f.}}
\newcommand{\cl}{c^{(\l)}}
%
\section{Introduction}

Recent experimental results for polarization fractions in
$\barB^{0,-} \to \phi(1020) \barKstar(892)^{0,-}$ are
\begin{eqnarray}
\begin{array}{lcl}
 f_L(\barB^0\to\phi\barKstarz) &=&
\begin{cases}
 0.506 \pm 0.040 \pm 0.015 & \text{\BABAR \cite{Aubert:2006uk}} \\
 0.45  \pm 0.05  \pm 0.02  & \text{\BELLE \cite{Chen:2005zv}} \\
 0.57  \pm 0.10  \pm 0.05  & \text{CDF \cite{Bussey2006}}
\end{cases}
,
\\
 f_\perp(\barB^0\to\phi\barKstarz) &=&
\begin{cases}
 0.227 \pm 0.038 \pm 0.013         & \text{\BABAR \cite{Aubert:2006uk}} \\
 0.31  ^{+0.06}_{-0.05}  \pm 0.02  & \text{\BELLE \cite{Chen:2005zv}} \\
 0.20  \pm 0.10  \pm 0.05          & \text{CDF \cite{Bussey2006}} \\
\end{cases}
,
\\
 f_L(B^- \to\phi K^{*-}) &=&
\begin{cases}
0.49\pm0.05\pm0.03 & \text{\BABAR \cite{:2007br}} \\
0.52\pm0.08\pm0.03 & \text{\BELLE \cite{Chen:2005zv}} \\
\end{cases},
\\
 f_\perp(B^-\to \phi K^{*-}) &=&
 \begin{cases}
 0.21  \pm 0.05  \pm 0.02    \quad \text{\BABAR \cite{:2007br}} \\
 0.19  \pm 0.08  \pm 0.02    \quad \text{\BELLE \cite{Chen:2005zv}}
\end{cases}.
\end{array}
\end{eqnarray}
Here, the polarization fractions $f_{\lambda}$ ($\lambda = L,\parallel,\perp$) are
given by
$f_\lambda =
|A_{\lambda}^2| / \sum_{\sigma=L,\parallel,\perp}|A_{\sigma}|^2 $,
with polarization amplitudes $A_{L}\equiv A_{0}$, $A_{\parallel}$ and
$A_{\perp}$ being longitudinal, parallel and perpendicular modes in the
transversity basis, respectively.
Experimental results show that $f_L \sim 0.5$ and $f_\perp \sim
f_\parallel$.
On the other hand, the power-counting estimate in the standard model
(SM) tells that the longitudinal mode is dominant \cite{Cheng:2001aa}.
In the SM, the QCD factorization (QCDF) calculation yields
\cite{Cheng:2001aa} $ f_L : f_\parallel : f_\perp =
1-\orderof{1/m_b^2}:\orderof{1/m_b^2}:\orderof{1/m_b^2} $.
%
The experimental results largely deviate from the intuition in the SM.
Similar discrepancies have been observed in penguin-dominated $B^{\pm,0}\to
\rho^{\pm,0} K^{*0}$ decays \cite{Abe:2004mq,Aubert:2006fs}.
These discrepancies are referred as the polarization puzzle/anomaly in
$B\to VV$ (where $V$ denotes a vector meson) decays.

Solutions to the puzzle have been discussed within or beyond the standard
model \cite{Cheng:2001aa,Li:2004ti,Li:2003he}.
The recipe of fine-tuning  form factors is proposed in \cite{Li:2004mp}.
Effects of final-state interactions are discussed in
\cite{Cheng:2004ru,Ladisa:2004bp}.
Sizable annihilation effects are considered
in \cite{Kagan:2004uw,Yang:2005tv,Beneke:2006hg}.
As discussed in \cite{Yang:2005tv}, the magnitude of annihilation
correction is of ${\cal O}\left[ 1/m_b^2 \log^2 m_b/\Lambda_h \right]$.
%
Furthermore, the effect is destructive to longitudinal, and constructive
to transverse modes.  Thus we may resolve the polarization puzzles by
introducing annihilation effects.
We note that, however, the perturbative QCD (pQCD) yields $f_L \gtrsim
0.75$ even with annihilation effects \cite{Li:2004ti}.

The $b\to s\g$ (where $\g$ denotes a gluon) operator,
which enhances the transverse components,
was discussed in \cite{Hou:2004vj}.
However, it was found \cite{Das-Yang,Kagan:2004uw}
that the contribution due to the
operator mainly affects the longitudinal mode.

As for the solutions of the puzzle,
the effects of NP-induced tensor operators
are discussed in \cite{Das-Yang}
and right-handed currents $\sbar\gamma^\mu(1+\gamma_5)
b\,\qbar\gamma_\mu(1\pm\gamma_5) q$ are in
\cite{Kagan:2004ia,Alvarez:2004ci,Chen:2006vs}.
Because the right-handed currents may decrease the magnitude
of $|A_0|$ and increase $|A_{\perp}|$,
it can explain the ratio $|A_\perp/A_0|$.
However, the resulting $|A_{\parallel}| \ll |A_{\perp}|$
\cite{Kagan:2004ia}
is in contrast with the data $|A_\parallel| \sim |A_\perp|$.

%
New physics (NP) contributions to $B\to\phi\Kstar$ decays due to $b\to
s\sbar s$ tensor operators, first mentioned in \cite{Kagan:2004uw},
are systematically discussed in \cite{Das-Yang}, and later the
idea is applied to $B\to \rho\Kstar$ by considering the 4-quark tensor
operators related to the processes $b\to s\dbar d$ and $s\ubar u$
\cite{Baek:2005jk}.
In the helicity basis,\footnote{Amplitudes in the helicity basis and the
transversity basis are related by
$\overline{A}_0=\overline{H}_{00}$,
$\overline{A}_\parallel = (\overline{H}_{++} + \overline{H}_{--})/\sqrt{2}$,
$\overline{A}_\perp = -(\overline{H}_{++} - \overline{H}_{--})/\sqrt{2}$.
}
four-quark tensor operators have leading effects to
$\overline{H}_{--}$ (or $\overline{H}_{++}$),
but sub-leading to $\overline{H}_{00}$.
The possibility of solving $B\to\phi\Kstar$ polarization puzzle by using
four-quark tensor operators is extensively studied in
\cite{Yang:2005tv,Das-Yang} and further investigated in \cite{Huang:2005qb,Yang:2004pm,Faessler:2007br,Chen:2005mka,Chang:2006dh}.
%

%
In \cite{Das-Yang} the general approach of resolving the
polarization anomaly of $B\to\phi\Kstar$
by using four-quark NP operators is studied.
There are two types of NP operators which are relevant to solve the
polarization anomaly.
They are tensor operators with
$\sigma_{\mu\nu}(1\pm\gamma_5)\otimes \sigma^{\mu\nu}(1\pm\gamma_5) $
structure.
The tensor operator
$\sigma_{\mu\nu}(1+\gamma_5)\otimes\sigma^{\mu\nu}(1+\gamma_5)$
results in $\overline{H}_{00}:\overline{H}_{--}:\overline{H}_{++} =
\orderof{1/m_b}:\orderof{1}:\orderof{1/m_b^2}$, while
$\sigma_{\mu\nu}(1-\gamma_5)\otimes \sigma^{\mu\nu}(1-\gamma_5)$ leads to
$\overline{H}_{00}:\overline{H}_{--}:\overline{H}_{++} =
 \orderof{1/m_b}:\orderof{1/m_b^2}:\orderof{1}$.

%
The decays $B\to PP$ (where $P$ denotes a pseudoscalar meson) are
sensitive to scalar/pseudoscalar 4-quark operators whereas
$B\to VV$ are sensitive to tensor operators.
Furthermore, it is known that scalar/pseudoscalar and tensor
operators are not independent; scalar/pseudoscalar operators can be
Fierz-transformed into tensor operators and vice versa.
Therefore, the combined analysis of scalar/pseudoscalar and
tensor operators for $B\to PP$ and $B \to VV$ modes will
give more severe constraints about NP scalar/pseudoscalar and
tensor operators.
%
%
\par
In this paper we focus on the $b\to s \bar s s$ decay processes. We consider the scalar/pseudoscalar operators induced by
Higgs penguin diagrams of the MSSM neutral Higgs bosons (NHB) \cite{Huang:2002ni,NHB,Cheng:2004jf,Huang:2005qb,Borzumati:1999qt},
while the tensor operators which are obtained from scalar/pseudoscalar
operators by the Fierz transformation.
In this NP scenario, such tensor operators, contribute to
the transverse polarization in $B\to\phi\Kstar$ decays, whereas original
scalar/pseudoscalar operators can affect $B \to K\etaP$ decays.
We study the consistency of both modes to see the
validity of the scenario. One should note that this NP effect is further suppressed by $m_q/m_s$ in the $b \to s \bar q q$ channel (with $q\equiv u$ or $d$) as compared with $b \to s \bar s s$. Although the recent observations of the sizable transverse fraction in $B^{\pm,0}\to
\rho^{\pm,0} K^{*0}$ decays may hint at large annihilation effects, the present study for decays $B\to\phi\Kstar$ and $B \to K\etaP$ can offer more severe constraints on the NP.
%

The organization of the present article is as follows:
In Sec.~\ref{sec-formulation}, we summarize the formulation of MSSM-NHB
scalar/pseudoscalar operators and its contributions to $B\to K\etaP$ and
$B\to \phi\Kstar$ decays.
In Sec.~\ref{sec-numerical}, we numerically analyze the decays for
$B\to K\etaP$ and $B\to\phi\Kstar$.
Sec.~\ref{sec-summary} is devoted to summary and discussions.
%
%
\section{Formulation}\label{sec-formulation}
\subsection{SM and NP operators}
In the SM the effective Hamiltonian relevant to
$B \to K \etaP$ and
$B \to \phi \Kstar$ decays is given by
\begin{eqnarray}
\calH_{\rm eff}^{\rm SM} &=& \frac{G_F}{\sqrt{2}}
\left[
 \sum_{q=u,c} V_{qb}V_{qs}^* (c_1(\mu) O_1^q(\mu) + c_2(\mu) O_2^q(\mu))
\right.
\nonumber\\&&
-
\left.
 V_{tb} V_{ts}^* \sum_{i=3}^{10} c_{i}(\mu) O_{i}(\mu)
 + c_{7\gamma}(\mu) O_{7\gamma}(\mu) + c_{8g}(\mu) O_{8g}(\mu)
\right] + \text{h.c.},
\end{eqnarray}
where the operators $O_{i=1,\ldots,10}$ are four-quark operators.
$O_{7\gamma}$ and $O_{8g}$ are electromagnetic and chromomagnetic dipole
operators, respectively. $\mu$ is the renormalization scale.
$V_{qb}$ and $V_{qs}$ ($q=u,c,t$) are elements of
 Cabibbo-Kobayashi-Maskawa (CKM) matrix.
The $b\to s\sbar s$ four-quark NP effective Hamiltonian is given by
\begin{eqnarray}
\calH_{\rm eff}^{\rm NP} =
-\frac{G_F}{\sqrt{2}}(V_{tb}V_{ts}^*) \sum_{i=11}^{26} c_i(\mu) O_i(\mu) + \text{h.c.},
\end{eqnarray}
where $O_i$ and $c_i$ ($i=11,\ldots,26$) are four-quark NP operators
introduced in \cite{Das-Yang}, and corresponding Wilson coefficients
\footnote{In \cite{Das-Yang}, CKM factors and Wilson coefficients
are not separated.}, respectively.
Explicit forms of $O_i$
($i=11,\ldots,26$) are shown in the following:
\\
(i) right-handed current operators
\begin{eqnarray}
\begin{array}{l}
O_{11}
= \sbar_\alpha \gamma_\mu \PR b_\alpha \, \sbar_\beta \gamma^\mu \PR s_\beta,
\quad
O_{12}
= \sbar_\alpha \gamma_\mu \PR b_\beta \, \sbar_\beta \gamma^\mu \PR s_\alpha,
\\
O_{13}
= \sbar_\alpha \gamma_\mu \PR b_\alpha \, \sbar_\beta \gamma^\mu \PL s_\beta,
\quad
O_{14}
= \sbar_\alpha \gamma_\mu \PR b_\beta \, \sbar_\beta \gamma^\mu \PL s_\alpha,
\end{array}
\end{eqnarray}
(ii) scalar/pseudoscalar operators
\begin{eqnarray}
\begin{array}{l}
O_{15}
= \sbar_\alpha \PR b_\alpha \, \sbar_\beta \PR s_\beta,
\quad
O_{16}
= \sbar_\alpha \PR b_\beta \, \sbar_\beta \PR s_\alpha,
\\
O_{17}
= \sbar_\alpha \PL b_\alpha \, \sbar_\beta \PL s_\beta,
\quad
O_{18}
= \sbar_\alpha \PL b_\beta \, \sbar_\beta \PL s_\alpha,
\\
O_{19}
= \sbar_\alpha \PR b_\alpha \, \sbar_\beta \PL s_\beta,
\quad
O_{20}
= \sbar_\alpha \PR b_\beta \, \sbar_\beta \PL s_\alpha,
\\
O_{21}
= \sbar_\alpha \PL b_\alpha \, \sbar_\beta \PR s_\beta,
\quad
O_{22}
= \sbar_\alpha \PL b_\beta \, \sbar_\beta \PR s_\alpha,
\end{array}
\end{eqnarray}
(iii) tensor/axial-tensor
operators\footnote{$\sigma_{\mu\nu}(1\pm\gamma_5)\sigma^{\mu\nu}(1\mp\gamma_5)$
type operators $O_{i}$ ($i=27,\dots,30$) in \cite{Das-Yang} are found to
vanish.}
\begin{eqnarray}
\begin{array}{l}
O_{23}
= \sbar_\alpha \sigma_{\mu\nu}\PR b_\alpha \,
    \sbar_\beta  \sigma^{\mu\nu}\PR s_\beta,
\quad
O_{24}
= \sbar_\alpha \sigma_{\mu\nu}\PR b_\beta  \,
    \sbar_\beta  \sigma^{\mu\nu}\PR s_\alpha,
\\
O_{25}
= \sbar_\alpha \sigma_{\mu\nu}\PL b_\alpha \,
    \sbar_\beta  \sigma^{\mu\nu}\PL s_\beta,
\quad
O_{26}
= \sbar_\alpha \sigma_{\mu\nu}\PL b_\beta  \,
    \sbar_\beta  \sigma^{\mu\nu}\PL s_\alpha.
\end{array}
\end{eqnarray}
Since $B\to PP$ ($B \to VV$) decays are not sensitive to the factorized
tensor (scalar/pseudoscalar) $b\to s\sbar s$ operators, it
is a good approximation to use $O_{i}$ with $i=1,\ldots,22$
($i=1,\ldots,14,23,\ldots,26$) for $B\to PP$ ($B\to VV$).
Some of these NP operators are not independent, and can be related with
each other by the Fierz transformation:
\begin{eqnarray}
\begin{array}{l}
O_{19} = -\tfrac{1}{2}O_{14},
\quad
O_{20} = -\tfrac{1}{2}O_{13},
\quad
O_{21} = -\tfrac{1}{2}O_{6},
\quad
O_{22} = -\tfrac{1}{2}O_{5},
\\
O_{23}  = -4 O_{15} -8 O_{16},
\quad
O_{24}  = -8 O_{15} -4 O_{16},
\\
O_{25}  = -4 O_{17} -8 O_{18},
\quad
O_{26}  = -8 O_{17} -4 O_{18}.
\end{array}
\label{Fierz-st}
\end{eqnarray}
Due to the Fierz transformation,
we can introduce modified Wilson coefficients $\bar{c}_{i}$,
which are defined by
\begin{eqnarray}
\bar{c}_{i} = c_{i} - \frac{1}{2} c_{j}
\quad
\text{with }  (i,j)=(5,22),\,(6,21),\,(13,20),\,(14,19),
\label{new-WC1}
\end{eqnarray}
and
\begin{eqnarray}
\begin{pmatrix}
\bar{c}_{15} \\ \bar{c}_{16} \\  \bar{c}_{23} \\ \bar{c}_{24}
\end{pmatrix}
=
M\begin{pmatrix}c_{15} \\ c_{16} \\ c_{23} \\ c_{24}\end{pmatrix},
\quad
\begin{pmatrix}
\bar{c}_{17} \\ \bar{c}_{18} \\  \bar{c}_{25} \\ \bar{c}_{26}
\end{pmatrix}
=
M\begin{pmatrix}c_{17} \\ c_{18} \\ c_{25} \\ c_{26}\end{pmatrix}
\text{ with }
M=
\begin{pmatrix}
1 & 0 & -4 & -8 \\
0 & 1 & -8 & -4 \\
\frac{1}{12} & -\frac{1}{6} & 1 & 0 \\
-\frac{1}{6} & \frac{1}{12} & 0 & 1
\end{pmatrix}.
\label{new-WC2}
\end{eqnarray}
Thus we can replace the Wilson coefficients by the effective ones:
\begin{eqnarray}
\{ c_i,\bar{c}_j \}
\quad i=1,\dots , 4,7,\dots , 12, 15, \dots ,18,
\quad j=5,6,13,14,
\quad
\label{replace-PP}
\end{eqnarray}
for $B\to PP$ decays,
and
\begin{eqnarray}
\{ c_i, \bar{c}_j \}
\quad i=1,\dots , 4, 7, \dots , 12,
\quad j=5,6,13,14,23,\dots ,26,
\label{replace-VV}
\end{eqnarray}
for $B\to VV$ decays,
so that the decay amplitudes can be simplified.
\subsection{$B\to K\etaP$ Decay Amplitudes}
In the SM, $B\to K\etaP$ decay amplitudes are given by \cite{Beneke:2003}
\begin{eqnarray}
A(\barB \to \barK \etaP) =
 \sum_{p=u,c} V_{pb}V_{ps}^* {\cal T}^p_{\barK\etaP},
\end{eqnarray}
where
\begin{eqnarray}
\sqrt{2}{\cal T}_{B\to K^-\etaP}^p
&=&
A_{\barK \etaP_q} \left[\delta_{pu} \alpha_2
                  + 2\alpha_3^p + \tfrac{1}{2} \alpha_{3,EW}^p
                  + 2\beta_{S3}^p \right]
\nonumber\\&&
+ \sqrt{2} A_{\barK\etaP_s} \bigl[
            \delta_{pu}\beta_{2}
            +\alpha_3^p + \alpha_4^p
            - \tfrac{1}{2}\alpha_{3,EW}^p
            - \tfrac{1}{2} \alpha_{4,EW}^p
            + \beta_3^p + \beta_{3,EW}^p + \beta_{S3}^p
\bigr]
\nonumber\\&&
+\sqrt{2} A_{\barK\etaP_c} \left[ \delta_{pc} \alpha_2 + \alpha_3^p \right]
\nonumber\\&&
+A_{\etaP_q \barK}\left[ \delta_{pu}(\alpha_1 + \beta_2) + \alpha_4^p +
          \alpha_{4,EW}^p + \beta_{3}^p + \beta_{3,EW}^p \right],
\label{eta-amp1}
\\
\sqrt{2}{\cal T}_{\barB^0 \to \barK^0 \etaP}^p &=&
A_{\barK \etaP_q} \left[ \delta_{pu} \alpha_2 + 2\alpha_3^p +
           \tfrac{1}{2}\alpha_{3,EW}^p  + 2\beta_{S3}^p \right]
\nonumber\\&&
+\sqrt{2} A_{\barK\etaP_s} \bigl[
                            \alpha_3^p + \alpha_4^p  -
                \tfrac{1}{2} \alpha_{3,EW}^p -\tfrac{1}{2}
                \alpha_{4,EW}^p
   + \beta_3^p - \tfrac{1}{2}\beta_{3,EW}^p + \beta_{S3}^p  \bigr]
\nonumber\\&&
+\sqrt{2} A_{\barK\etaP_c} \left[\delta_{pc} \alpha_2 +
                \alpha_3^p\right]
\nonumber\\&&
+ A_{\etaP_q \barK} \left[ \alpha_4^p - \tfrac{1}{2} \alpha_{4,EW}^p +
             \beta_3^p - \tfrac{1}{2} \beta_{3,EW}^p \right].
\label{eta-amp2}
\end{eqnarray}
For the $B\to PP$ decays $\alpha_{1,2,3,4,3EW,4EW}$
are defined  as
\begin{eqnarray}
\begin{array}{c}
\alpha_{1,2} = a_{1,2},\quad
\alpha_{3}^p = a_3^p - a_5^p,\quad
\alpha_{4}^p = a_4^p + r_\chi^{M_2} a_6^p,
\\
\alpha_{3,EW}^p = a_9^p - a_7^p,
\quad
\alpha_{4,EW}^p = a_{10}^p + r_\chi^{M_2} a_8^p,
\end{array}
\end{eqnarray}
with $r_\chi^{K} = 2m_K^2 / m_b(m_q + m_s)$ and
 $r_\chi^{\etaP_s} \equiv h^s_{\etaP} / (f^s_{\etaP}m_b m_s)$.
$A_{M_1M_2}$ are given by
\begin{eqnarray}
A_{M_1M_2} = i \frac{G_F}{\sqrt{2}}\cdot m_B^2 F_0^{B\to M_1}(0) f_{M_2}.
\end{eqnarray}
Contributions from annihilation diagram are represented by $\beta_Q$
($Q=3,4,3EW,4EW,S3$), where $\beta_Q \equiv b_Q \cdot
B_{M_1M_2}/A_{M_1M_2}$ with
\begin{eqnarray}
B_{K\etaP_r}= i \frac{G_F}{\sqrt{2}} f_B f_K f_{\etaP}^r,
\quad
B_{\etaP_q K} = i \frac{G_F}{\sqrt{2}} f_B f_K
f_{\etaP}^q,
\end{eqnarray}
(where $r = q$ or $s$) respectively.
$b_{3,4,3EW,4EW}$ are the coefficients due to
weak annihilation of penguin operators.
$b_{S3}$ is originated from the singlet penguin contribution
which is introduced in \cite{Beneke:2002,Beneke:2003}.
Note that following the approximation adopted in \cite{Beneke:2003}, we have neglected single weak annihilations $\beta_{S1}$, $\beta_{S2}$, $\beta_{S3,EW}$, and only keep $\beta_{S3}$.

In the above results, we adopt $a_{i=1,...,10}^p$ and $b_Q^p$, given by
the QCD factorization (QCDF) calculation in \cite{Beneke:2003}. The NP effects due to scalar/pseudoscalar operators can be
included by replacing $c_{5}$ and $c_{6}$ with the effective ones
$c_{5(6)}^{\rm eff} \equiv c_{5(6)} + \Delta c_{5(6)}$ in the $\bar B\to \bar K \eta_s$ decay amplitudes in the following way:
\begin{eqnarray}
\begin{array}{l}
\Delta c_5 = \tfrac{1}{2}(-\bar{c}_{16} + \bar{c}_{18} + c_{20} - c_{22}),
\quad
\Delta c_6 =  \tfrac{1}{2}(-\bar{c}_{15} + \bar{c}_{17} + c_{19} - c_{21}), ~~~{\rm for}~~ \alpha_4^p,\ \beta_3^p\,,
\end{array}
\label{replace-43}
\end{eqnarray}
and \begin{eqnarray}
\begin{array}{l}
\Delta c_5 = \tfrac{1}{2}(c_{20} - c_{22}),
\quad
\Delta c_6 =  \tfrac{1}{2}(c_{19} - c_{21}), ~~~{\rm for}~~
 \alpha_3^p,\ \beta_2^p,\ \beta_{S3}\,.
\end{array}
\label{replace-323}
\end{eqnarray}
Here $\bar{c}_{15}$, $\bar{c}_{16}$, $\bar{c}_{17}$, and $\bar{c}_{18}$
are defined in (\ref{new-WC2}).
$\eta$ and $\eta'$ mesons states can be regarded as mixed states of
$\ket{\eta_q} \equiv \frac{1}{\sqrt{2}}(\ket{\ubar u} + \ket{\dbar d})$
 and $\ket{\eta_s} \equiv \ket{\sbar s}$ with a mixing angle
$\phi_\eta$ \cite{Beneke:2002}:
\begin{eqnarray}
\begin{pmatrix} \ket{\eta} \\ \ket{\eta'} \end{pmatrix}
=
\begin{pmatrix}
\cos \phi_\eta & -\sin\phi_\eta \\
\sin \phi_\eta & \phantom{-} \cos\phi_\eta
\end{pmatrix}
\begin{pmatrix} \ket{\eta_q} \\ \ket{\eta_s} \end{pmatrix}.
\label{eta-mix}
\end{eqnarray}
Decay constants $f^{q,s}_{\etaP}$,
pseudoscalar densities $h^{q,s}_{\etaP}$ are defined by
\begin{eqnarray}
\begin{array}{l}
\displaystyle
\bra{\etaP(p)}\qbar\gamma^\mu\gamma_5 q\ket{0} =
 -\frac{i}{\sqrt{2}}f^q_{\etaP} p^\mu,
\quad
\bra{\etaP(p)}\sbar\gamma^\mu\gamma_5 s\ket{0} =
 -i f^s_{\etaP} p^\mu,
\\
\displaystyle
2m_q \bra{\etaP(p)}\qbar \gamma_5 q \ket{0}
 = - \frac{i}{\sqrt{2}}h^q_{\etaP},
\quad
2m_s \bra{\etaP(p)}\qbar \gamma_5 q \ket{0}
 = - i h^s_{\etaP},
\end{array}
\end{eqnarray}
with $m_q = (m_u+m_d)/2$.
As for explicit forms of $f_{\etaP}^{q,s}$, $h_{\etaP}^{q,s}$
and form factors $F_0^{B\to \etaP}(q^2)$,
we summarize in Appendix \ref{apdx-etap}.
%


\subsection{$B\to \phi\Kstar$ Decay Amplitudes}
Decay amplitudes of $\barB \to \phi \barKstar$
can be decomposed as
\begin{eqnarray}
A(\barB \to \phi \barKstar)
=
\sum_{h=0,\pm} \overline{H}_{hh},
\end{eqnarray}
where
\begin{eqnarray}
\overline{H}_{hh} =
\sum_{p=u,c} V_{pb}V_{ps}^*
\left( {\cal T}_{\phi\Kstar,A}^{p,h} + {\cal T}_{\phi\Kstar,B}^{p,h}
\right),
\quad (h=0,\pm),
\end{eqnarray}
is the amplitudes in the helicity basis.
Amplitudes for the emission topology (the ${\cal T}_A$ part)
is given by\footnote{The coefficient ``1/2'' in front of $a^h_{24,26}$
can be realized as follows. Take $O_{23}$ as an example.
Because, under the Fierz transform, $O_{23}$ can be written by
\begin{eqnarray}
 O_{23}= (1/2)\sbar_\alpha \sigma_{\mu\nu}\PR b_\beta \, \sbar_\beta  \sigma^{\mu\nu}\PR s_\alpha - 6 \sbar_\alpha \PR b_\beta \, \sbar_\beta  \PR s_\alpha \,, \label{footnote:ai}
\end{eqnarray}
therefore, in the factorization limit, we obtain \begin{eqnarray}
 \langle \phi \bar K^*|O_{23}|\bar B\rangle
 &=&\langle \phi|\sbar \sigma^{\mu\nu}\PR s|0\rangle \langle \bar K^*|\sbar \sigma_{\mu\nu}\PR b|\bar B\rangle\nonumber\\
 & & + \tfrac{1}{2}\langle \phi|\sbar_\beta \sigma^{\mu\nu}\PR s_\alpha|0\rangle \langle \bar K^*|\sbar_\alpha \sigma_{\mu\nu}\PR b_\beta|\bar B\rangle \nonumber \\
 &=& \Bigg(1 +\frac{1}{2N_c}\Bigg)\langle \phi|\sbar \sigma^{\mu\nu}\PR s|0\rangle \langle \bar K^*|\sbar \sigma_{\mu\nu}\PR b|\bar B\rangle.
 \end{eqnarray}
Note that the second term in the right hand side of (\ref{footnote:ai}) gives no contribution since the local scalar current cannot couple to $\phi$. Similarly, in the factorization limit we have
\begin{eqnarray}
 \langle \phi \bar K^*|O_{24}|\bar B\rangle
 &=& \Bigg(\frac{1}{N_c} +\frac{1}{2} \Bigg)\langle \phi|\sbar \sigma^{\mu\nu}\PR s|0\rangle \langle \bar K^*|\sbar \sigma_{\mu\nu}\PR b|\bar B\rangle.
 \end{eqnarray}
The same procedure can be applied to the matrix elements containing $O_{25}$ and $O_{26}$.}
\begin{eqnarray}
\sum_{p=u,c} V_{pb}V_{ps}^*
{\cal T}_{\phi\Kstar,A}^{p,h}
&=&
(-V_{tb}V_{ts}^*)
\left\{
A^h_{(\barB\barKstar,\phi)-}
\left[ a_{3}^h + a_{4}^h + a_{5}^h - r_{\chi}^\phi a_6^h
      -\tfrac{1}{2}
       \left(
        a_{7}^h - r_{\chi}^\phi a_8^h + a_{9}^h + a_{10}^h
       \right)
\right]
\right.
\nonumber\\&&
+A^h_{(\barB\barKstar,\phi)+}
\left[
  a_{11}^h + a_{12}^h + a_{13}^h - r_{\chi}^\phi a_{14}^h
\right]
\nonumber\\&&
\left.
+ A^h_{(\barB\barKstar,\phi)T+}
\left[ a_{23}^h + \tfrac{1}{2} a_{24}^h \right]
+ A^h_{(\barB\barKstar,\phi)T-}
\left[ a_{25}^h + \tfrac{1}{2} a_{26}^h \right]
\right\},
\label{phiKstar-amp}
\end{eqnarray}
with $r_\chi^\phi$ given by
\begin{eqnarray}
r_\chi^\phi = \frac{2m_\phi}{m_b(\mu)}\frac{f_\phi^T(\mu)}{f_\phi}.
\end{eqnarray}
Coefficients $a_{i}^{h}$ ($i=3,\dots,10$) have been calculated in
QCDF \cite{Beneke:2003,Beneke:2006hg}.
However instead of $c_5$ and $c_6$, $\bar{c}_5$ and $\bar{c}_6$ should be used in the calculation of $a_5$ and $a_6$ (see (\ref{replace-VV})).
\begin{eqnarray}
a_i^h = \left(\bar{c}_i +\frac{\bar{c}_{i\pm1}}{N_c}\right)
  + {\cal O}(\alpha_s),
\text{ with } i=23,24,25,26,
\end{eqnarray}
where the radiative corrections are negligible.
We summarized the explicit form of $a_{i}^{h}$ for
$i=11,\dots,14$ due to the right-handed four-quark operators
 in Appendix \ref{app:ai_bi}.
%

In (\ref{phiKstar-amp}), coefficients
$A_{(BV_1,V_2)\pm}^h$ and $A_{(BV_1,V_2),T\pm}^h$ are given by
\begin{eqnarray}
A_{(\barB\barKstar,\phi)\mp}^h
&\equiv&
\frac{G_F}{\sqrt{2}}
 \bra{\phi(q,\varepsilon_1(h))}\sbar\gamma^\mu(1-\gamma_5) s \ket{0}
 \bra{\barKstar(p',\varepsilon_2(h))}\sbar\gamma_\mu(1\mp\gamma_5) b \ket{\barB(p)}
\nonumber\\
&=&
\frac{G_F}{\sqrt{2}}
\{
i f_\phi m_{\phi}
\left[\frac{-2i}{m_B+m_{\Kstar}}\epsilon_{\mu\nu\alpha\beta}
\varepsilon_{1}^{\mu*} \varepsilon_{2}^{\nu*} p^\alpha p'^\beta V(q^2)
\right]
\nonumber\\&&
\mp i f_\phi m_\phi \left[
(m_B + m_{\Kstar})(\varepsilon_{1}^* \cdot\varepsilon_{2}^*) A_1(q^2)
- (\varepsilon_{1}^* \cdot p)(\varepsilon_2^* \cdot p)
\frac{2A_2(q^2)}{m_B + m_{\Kstar}}
\right]
\},
\\
A_{(\barB\barKstar,\phi),T\pm}^h
&\equiv&
\frac{G_F}{\sqrt{2}}
 \bra{\phi(q,\varepsilon_1(h))}
    \sbar \sigma_{\mu\nu} s (1\pm\gamma_5)\ket{0}
 \bra{\barKstar(p',\varepsilon_2(h))}
    \sbar\sigma^{\mu\nu}(1\pm\gamma_5) b\ket{\barB(p)}
\nonumber\\
&=&
 \frac{G_F}{\sqrt{2}} f_\phi^T
\{
 8\epsilon_{\mu\nu\rho\sigma} \varepsilon_1^{\mu*}
 \varepsilon_2^{\nu2}p^\rho p'^\sigma T_1(q^2)
\nonumber\\&&
\mp 4iT_2(q^2)
 \left[(\varepsilon_1^{*}\cdot
        \varepsilon_2^*)(m_B^2-m_{\Kstar}^2)
   - 2(\varepsilon_1^* \cdot p)(\varepsilon_2^* \cdot p)
 \right]
\nonumber\\&&
\pm 8iT_3(q^2) (\varepsilon_1^* \cdot p)(\varepsilon_2^* \cdot p)
\frac{m_{\phi}^2}{m_B^2 - m_{\Kstar}^2}
\},
\end{eqnarray}
or in the explicit forms
\begin{eqnarray}
\begin{array}{lcl}
A_{(\barB\barKstar,\phi)\mp}^0   &=&
\displaystyle
 \mp \frac{G_F}{\sqrt{2}} (if_\phi m_\phi)
 (m_B + m_{\Kstar})[a A_1(m_\phi^2) - b A_2(m_\phi^2)],
\\
A_{(\barB\barKstar,\phi)-}^{\pm} &=&
\displaystyle
 \phantom{-}\frac{G_F}{\sqrt{2}}(if_\phi m_\phi)
 \left[(m_B + m_{\Kstar}) A_1(m_\phi^2)
   \mp \frac{2m_B p_c}{m_B + m_{\Kstar}} V(m_\phi^2)\right],
\\
A_{(\barB\barKstar,\phi)+}^\pm &=&
\displaystyle
 \phantom{-}\frac{G_F}{\sqrt{2}}(if_\phi m_\phi)
 \left[-(m_B + m_{\Kstar}) A_1(m_\phi^2)
   \mp \frac{2m_B p_c}{m_B + m_{\Kstar}} V(m_\phi^2)\right],
\\
A_{(\barB\barKstar,\phi)T\pm}^0   &=&
\displaystyle
 \mp \frac{G_F}{\sqrt{2}}4(if_\phi^T) m_B^2
 [ h_2 T_2(m_\phi^2) - h_3 T_3(m_\phi^2)],
\\
A_{(\barB\barKstar,\phi)T+}^\pm &=&
\displaystyle
 -\frac{G_F}{\sqrt{2}}
4(if_\phi^T) m_B^2 [ \pm f_1 T_1(m_\phi^2) - f_2 T_2(m_\phi^2)] ,
\\
A_{(\barB\barKstar,\phi)T-}^\pm &=&
\displaystyle
 -\frac{G_F}{\sqrt{2}}
4(if_\phi^T) m_B^2 [\pm f_1 T_1(m_\phi^2) + f_2 T_2(m_\phi^2)] ,
\end{array}
\end{eqnarray}
with $a = (m_B^2 - m_{\phi}^2 - m_{\Kstar}^2)/(2m_{\phi}m_{\Kstar})$,
$b=(2m_B^2 p_c^2)/[m_\phi m_{\Kstar}(m_B + m_{\Kstar})]$
and
\begin{eqnarray}
\begin{array}{l}
f_1 = 2p_c/m_B,
\quad
f_2 = (m_B^2 - m_{\Kstar}^2)/m_B^2,
\\
\displaystyle h_2 = \frac{1}{2m_{\Kstar}m_{\phi}}
\left[
\frac{(m_B^2 - m_\phi^2 - m_{\Kstar}^2)(m_B^2 - m_{\Kstar}^2)}{m_B^2}- 4p_c^2
\right],
\\
\displaystyle
h_3 = \frac{1}{2m_{\Kstar}m_\phi}
\left(\frac{4p_c^2 m_\phi^2}{m_B^2 - m_{\Kstar}^2}\right).
\end{array}
\label{a2325}
\end{eqnarray}
Here we have used decay constants and form factors defined in Appendix
\ref{decay-form}.

Weak annihilation contributions
 (the ${\cal T}_B$-part) to $\barB^0 \to \phi
\barK^{*0}$ and $B^- \to \phi K^{*-}$ decay amplitudes in helicity basis can be given by
\begin{eqnarray}
\begin{array}{lcl}
\displaystyle \sum_{p=u,c} V_{pb}V_{ps}^* {\cal
T}_{\phi\Kstarz,B}^{p,h} &=& \displaystyle B_{\phi\Kstar}
(-V_{tb}V_{ts}^*) \left[ b_{3}^h - \tfrac{1}{2} b_{3EW}^h +
b_5^h\right],
\\
\displaystyle \sum_{p=u,c} V_{pb} V_{ps}^* {\cal T}_{\phi
K^{*-},B}^{p,h} &=& \displaystyle B_{\phi\Kstar} \left\{
(-V_{tb}V_{ts}^*) \left[ b_{3}^h + b_{3EW}^h + b_5^h
\right] + V_{ub}V_{us}^* \cdot b_2^h \right\},
\end{array}
\label{phiKstar-ann-coeff}
\end{eqnarray}
where
\begin{eqnarray}
B_{\phi\Kstar} \equiv i \frac{G_F}{\sqrt{2}}f_B f_{\Kstar} f_\phi,
\end{eqnarray}
and
\begin{eqnarray}
\begin{array}{lcl}
b_3^h &=&
\displaystyle\frac{C_F}{N_c^2}
\left[ c_3 A_1^{i,h} + \bar{c}_5(A_3^{i,h} + A_3^{f,h})
                     + N_c \bar{c}_6 A_3^{f,h}\right],
\\
b_{3EW}^h &=&
\displaystyle \frac{C_F}{N_c^2}
\left[ c_9 A_1^{i,h} + c_7(A_3^{i,h} + A_3^{f,h}) + N_c c_8
 A_3^{f,h}\right],
\\
b_{2}^{h} &=& \displaystyle \frac{C_F}{N_c^2} c_2 A_1^{i,h},
\\
b_5^h &=& \displaystyle - \frac{C_F}{N_c^2} \left[ c_{11} A_2^{i,h}
+ \bar{c}_{13} (A_3^{i,h} + A_3^{f,h}) +  N_c \bar{c}_{14} A_3^{f,h}\right],
\end{array}
\end{eqnarray}
with $h=0,-,+$.
Building blocks $A_{1,3}^{i(f),h}$ can be found in Appendix of
\cite{Beneke:2006hg}.
%

\subsection{Scalar and Pseudoscalar operators in the MSSM}
In the MSSM, scalar and pseudoscalar operators can be induced by
neutral-Higgs boson (NHB) penguin diagrams. We refer such operator as
the MSSM-NHB scalar/pseudoscalar operators.
In \cite{Huang:2002ni}, $b\to s \lbar\l$ (where $\l$ denotes a charged
lepton) scalar/pseudoscalar operators $Q_{1,2}^{(\prime)}$ induced by
the MSSM-NHB penguin diagrams are considered.
%
$b\to s \qbar q$ scalar/pseudoscalar operators can be obtained by
replacing Higgs-$\lbar$-$\l$ vertex with Higgs-$\qbar$-$q$ vertex
\cite{Cheng:2004jf}. They are
\begin{eqnarray}
\begin{array}{ll}
\displaystyle
{\cal O}_{15}  = \sbar(1+\gamma_5) b
                 \sum_q \frac{m_q}{m_b} \qbar(1+\gamma_5) q,
&\displaystyle
{\cal O}_{16} = \sbar_i(1+\gamma_5) b_j
                \sum_q \frac{m_q}{m_b}\qbar_j(1+\gamma_5) q_i,
\\ \displaystyle
{\cal O}_{17} = \sbar(1-\gamma_5)b
                \sum_q \frac{m_q}{m_b}\qbar(1-\gamma_5) q,
& \displaystyle
{\cal O}_{18} = \sbar_i(1-\gamma_5) b_j
                \sum_q \frac{m_q}{m_b}\qbar_j(1-\gamma_5) q_i,
\\ \displaystyle
{\cal O}_{19} = \sbar(1+\gamma_5) b
                \sum_q \frac{m_q}{m_b} \qbar(1-\gamma_5) q,
& \displaystyle
{\cal O}_{20} = \sbar_i(1+\gamma_5) b_j
                \sum_q \frac{m_q}{m_b}\qbar_j(1-\gamma_5) q_i,
\\ \displaystyle
{\cal O}_{21} = \sbar(1-\gamma_5) b
                \sum_q \frac{m_q}{m_b}\qbar(1+\gamma_5) q,
& \displaystyle
{\cal O}_{22} = \sbar_i(1-\gamma_5) b_j
                \sum_q \frac{m_q}{m_b}\qbar_j(1+\gamma_5) q_i,
\end{array}
\end{eqnarray}
where $q=u,d,s,c$.\footnote{Precisely speaking, in the two doublet Higgs
model, couplings of the light neutral Higgs $h^0$ to up-type quarks are
suppressed by $\tan\beta$ compared with the down-type quarks. Therefore
we can neglect contributions of  up-type quarks. }
The Wilson coefficients ${\cal C}_{i}(\mu)$ of ${\cal
O}_{i}$ with $i=15,\ldots,22$ at $\mu=m_W$ are given by
\cite{Huang:2002ni,Cheng:2004jf}:
\begin{eqnarray}
{\cal C}_{15}(m_W)  = \frac{e^2}{16\pi^2} (C_{S}  + C_{P}),
\quad
{\cal C}_{17}(m_W)
 = \frac{e^2}{16\pi^2} (C_{S}' - C_{P}'),
\nonumber
\\
{\cal C}_{19}(m_W)  = \frac{e^2}{16\pi^2} (C_{S}  - C_{P}),
\quad
{\cal C}_{21}(m_W)
 = \frac{e^2}{16\pi^2}(C_{S}' + C_{P}'),
\\
{\cal C}_{i} = 0  \quad (i=16,18,20,22),
\nonumber
\end{eqnarray}
and four-quark tensor operators are not directly induced.
Here
\begin{eqnarray}
\begin{array}{lcl}
\displaystyle
C_{S}^{(\prime)}
&=&
\displaystyle \frac{4}{3\lambda_t}\frac{g_s^2}{g_2 \sin^2\theta_W}
\frac{m_b^2}{m_{H^0}^2}
\frac{\cos\alpha^2 + r_s \sin^2\alpha}{\cos^2\beta}
\frac{m_{\tilde{g}}}{m_b}
f_b'(x) \delta_{23}^{dLL(RR)} \delta_{33}^{dLR(LR*)},
\\
\displaystyle
C_{P}^{(\prime)}
&=&
\displaystyle\mp
\frac{4}{3\lambda_t}\frac{g_s^2}{g_2 \sin^2\theta_W}
\frac{m_b^2}{m_{A^0}^2}
(r_p + \tan^2\beta)
\frac{m_{\tilde{g}}}{m_b}
f_b'(x) \delta_{23}^{dLL(RR)} \delta_{33}^{dLR(LR*)},
\end{array}
\label{mssm-coeff-C}
\end{eqnarray}
with $r_s = m_{H^0}^2/m_{h^0}^2$, $r_p = m_{A^0}^2/m_{Z^0}^2$,
$x=m_{\tilde{q}}^2/m_{\tilde{g}}^2$ and
$\lambda_t \equiv V_{tb}V_{ts}^*$.
$g_2$ and $g_s$ are gauge couplings
for the weak and strong interactions, respectively.
$m_{h^0}$, $m_{H^0}$ and $m_{A^0}$ are masses of neutral Higgs bosons
$h^0$, $H^0$, $A^0$, respectively.
$\alpha$ is the neutral Higgs mixing angle and $\tan\beta$ is the ratio
of the two Higgs vacuum expectation values,
$m_{\tilde{g}}$ and $m_{\tilde{q}}$ are the gluino mass and common
squark mass, respectively.
Factors $\delta_{23}^{dLL}$, $\delta_{23}^{dRR}$ and $\delta_{33}^{dLR}$
are down-type left-light second-third, right-right second-third and
left-right third generation squark mixing parameters, respectively.
The loop function $f_b'(x)$ is defined as $f_b'(x) \equiv
(x^2/2)\partial^2 f_{b0}(x)/\partial x^2$ and $f_{b0}(x)$ is defined in
\cite{Cheng:2004jf}. $f_b'(x)$ is given by
\begin{eqnarray}
f_b'(x) = - \frac{x(-1+x^2-2x\log x)}{2(x-1)^3}.
\end{eqnarray}

Because $O_{i}$ ($i=15,...,22$) and ${\cal O}_{i}$ are related by
\begin{eqnarray}
\frac{m_s}{m_b} O_{i} \subseteq {\cal O}_{i},
\end{eqnarray}
the Wilson coefficients $c_i(\mu)$ for $O_i(\mu)$ with
 $i=15,\ldots,26$ at $\mu = m_W$ are given by
\begin{eqnarray}
c_{15}(m_W) = \frac{m_s}{m_b} {\cal C}_{15} = D(A-B)\xi , \quad
c_{17}(m_W) = \frac{m_s}{m_b} {\cal C}_{17} = D(A-B)\xi',
\nonumber
\\
c_{19}(m_W) = \frac{m_s}{m_b} {\cal C}_{19} = D(A+B)\xi , \quad
c_{21}(m_W) = \frac{m_s}{m_b} {\cal C}_{21} = D(A+B)\xi', \\
\displaystyle
c_i=0 \quad \text{ for } i=16,18,20,22,23,24,25,26,
\nonumber
\label{MSSM-Wilson-coeff}
\end{eqnarray}
where
\begin{eqnarray}
\begin{array}{l}
D \equiv \dfrac{1}{12\pi^2}
\dfrac{1}{\lambda_t}
\dfrac{e^2 g_s^2}{g^2 \sin^2\theta_W}
 f_b'(m_{\tilde{q}}^2 / m_{\tilde{g}}^2 ) m_s m_{\tilde{g}},
\\
A \equiv
\dfrac{1}{m_{H^0}^2}
\left(\dfrac{\cos^2\alpha+(m_{H^0}^2/m_{h^0}^2)\sin^2\alpha
}{
\cos^2\beta}\right),
\quad
B \equiv \dfrac{1}{m_{A^0}^2}
\left(\dfrac{m_{A^0}^2}{m_{Z^0}^2} + \tan^2\beta \right),
\\
\xi  \equiv \delta_{23}^{dLL}\delta_{33}^{dLR},
\quad
\xi' \equiv \delta_{23}^{dRR}\delta_{33}^{dLR*}.
\end{array}
\label{MSSM-param}
\end{eqnarray}
Since the Wilson coefficients for $(\sbar b)(\dbar d)$ scalar and
pseudoscalar operators are suppressed by
$m_d/m_s$, we neglect contributions for operators with $q=d$ and we
consider only $b\to s\sbar s$ type operators.
In (\ref{MSSM-param}),
we note that $D$ is almost positive and real because $f_b'(x) < 0$,
$\Re\lambda_t<0$ and the imaginary part of $\lambda_t$
is negligibly small.

There are three enhancement factors: $\tan\beta$, $1/m_{A^0}$ and
$1/m_{H^0}$ in Eq. (\ref{MSSM-param}).
Therefore, if $\delta_{33}^{dLR}$ is sizable and neutral Higgses, $A^0$
and $H^0$, are sufficiently light and $\tan\beta$ is large, then the
Wilson coefficients for scalar/pseudoscalar operators can be large
enough.
In the present paper, for simplicity, we neglect the mixing between
scalar/pseudoscalar and tensor operators through
renormalization-group equations (RGEs) since such a mixing effect is
small.
Thus we assume $ c_{i}(m_b) \propto c_{i}(m_W)$ and we use same symbols
$A$, $B$, $D$, $\xi$ and $\xi'$ at $\mu \sim m_b$.

Under the existence of the MSSM-NHB scalar/pseudoscalar operators
(\ref{MSSM-param}), $\Delta c_{5,6}$, given in  (\ref{replace-43}), (\ref{replace-323}), in $B\to K\eta_s$ decays are rewritten
by
\begin{eqnarray}
\Delta c_6 &=&
\begin{cases}
 DB(|\xi|e^{\pm i\phi} - |\xi'|e^{\pm i\phi'}),
  &\text{ for } \alpha_4,\beta_3,
  \\
 \displaystyle
 \frac{1}{2}DB \left(2 - \frac{B-A}{B}\right)
 (|\xi|e^{\pm i\phi} - |\xi'|e^{\pm i\phi'}),
  &\text{ for } \alpha_3,\beta_2,\beta_{S 3},
 \\
 \end{cases}
\nonumber
\\
\Delta c_5 &=& 0,
\end{eqnarray}
where $\xi^{(\prime)} = |\xi^{(\prime)}| \exp(i\phi^{(\prime)})$.
%
%
As for the Wilson coefficients in $B\to\phi\Kstar$ decays,
we use the replacements (\ref{replace-VV}).
They are given by
\begin{eqnarray}
\begin{array}{l}
\displaystyle
\bar{c}_6 - c_6 = - \frac{1}{2} D (A+B) \xi'
= \frac{1}{2} \left( \frac{B-A}{B} - 2\right) DB |\xi'| e^{i\phi'},
\\
\displaystyle
\bar{c}_{14} = - \frac{1}{2} D (A+B) \xi
= \frac{1}{2} \left( \frac{B-A}{B} - 2\right) DB |\xi| e^{i\phi},
\\
\bar{c}_5 = c_{5},
\quad
c_{11} = c_{12} = \bar{c}_{13} = 0,
\\
\bar{c}_{23} = \frac{1}{12}D(A-B)\xi,
\quad
\bar{c}_{24} = -\frac{1}{6}D(A-B)\xi,
\\
\bar{c}_{25} = \frac{1}{12}D(A-B)\xi',
\quad
\bar{c}_{26} = -\frac{1}{6}D(A-B)\xi'.
\end{array}
\label{KetaP-effect}
\end{eqnarray}
As for $a_{23-26}$, 
we parametrized the following coefficients that appear in (\ref{phiKstar-amp}).
\begin{eqnarray}
\begin{array}{lcl}
\displaystyle
a_{23} + \frac{1}{2} a_{24}
\approx
\dfrac{1}{8N_c}
\dfrac{B-A}{B} DB |\xi| e^{i(\delta\pm\phi)},
\\
\displaystyle
a_{25} + \frac{1}{2} a_{26}
\approx
\dfrac{1}{8N_c}
\dfrac{B-A}{B} DB|\xi'| e^{i(\delta'\pm\phi')},
\end{array}
\label{phiKstar-MSSM}
\end{eqnarray}
where the $\alpha_s$ corrections are negligible. However, we still parametrize the strong phases $\delta$ and $\delta'$ here. Actually the strong phases consistent with zero in the fit. Here and below the helicity labels are omitted for $a_{23-26}$ since these coefficients very weakly depend on their helicities.

\begin{figure}[tbp]
\caption{$\theratio$ for various $\tan\beta$ and $m_A$.}\label{abratio}
\includegraphics[width=4in]{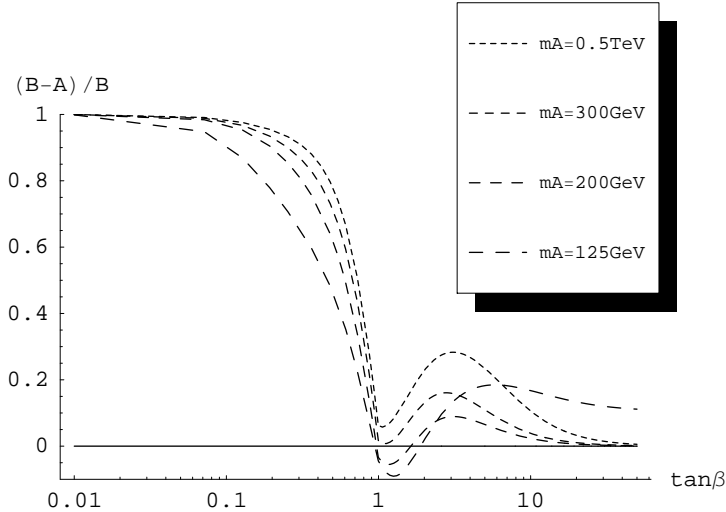}
\end{figure}

The factor $\theratio$ depends on the details of neutral Higgs
sector.
In FIG.~\ref{abratio}, we have plotted the value of $\theratio$ for
various $m_A$ and $\tan\beta$ in the MSSM.
In the MSSM, $\theratio$ is always smaller than one and $-0.1 \lesssim \theratio
\lesssim 0.3$ for $\tan\beta \gtrsim 1$.
When $\theratio \sim 0.2$, the ratio $|(a_{23(25)}+\frac{1}{2}a_{24(26)})/\Delta c_6|
\approx 0.02 \theratio \approx \orderof{10^{-2}}$.
%

\section{Numerical Analysis}\label{sec-numerical}
%
\subsection{Numerical Inputs}
We summarize input parameters in Table~\ref{inputs}.
As for $B\to \Kstar$ vector and tensor form factors, we follow the light-cone sum-rule (LCSR) results \cite{LCSR}, defined as
\begin{eqnarray}
F(q^2) = F(0) \exp (c_1 q^2/m_B^2 + c_2 q^2/m_B^2),
\end{eqnarray}
for $F\equiv A_0,\, A_1,\, A_2,\, V,\, T_1,\, T_2,\, T_3$,
where $c_1$ and $c_2$ are listed in Table~III of
\cite{Das-Yang}.
We use
$m_{B^\pm} \approx m_{B^0}=5.279\GeV$.
We parametrize $\lambda_B$ to be
$m_B/\lambda_B \equiv \int_0^1 dy \Phi_{B1}(y)/y$, where $\Phi_{B1}(y)$ is one of the two $B$ meson light
cone distribution amplitudes
with $y$ being the momentum fraction carried by
the light spectator quark in the $B$ meson.
As for the renormalization scale we use $\mu = m_b/2$.
In the calculation of the hard spectator and the weak annihilation,
we adopt $\mu_h = \sqrt{\Lambda_h\cdot\mu} \sim 1\GeV$
(with $\Lambda_h = 0.5\GeV$) corresponding to the hadronic scale.
\begin{table}[tbp]
\caption{Input Parameters}\label{inputs}
\begin{ruledtabular}
\begin{tabular}{ll}
Decay constants \cite{Cheng:2001aa,Beneke:2003} & $f_{\pi}=131\MeV$ \quad
                  $f_{K}=160\MeV$ \quad
                  $f_B= 210\MeV$ \quad
\\
                & $f_q = (1.07\pm0.02) f_\pi$ \quad
                  $f_s = (1.34\pm0.06) f_\pi$ \quad

\\
                & $f_{\Kstar} = 218\MeV$ \quad
                  $f_{\Kstar}^T = 175\MeV$ \quad
                  $f_\phi = 221\MeV$ \quad
                  $f_\phi^T = 175\MeV $
\\
$B$ meson parameter \cite{Beneke:2003}
                     &  $\lambda_B = 200_{-0}^{+250} \MeV$
\\
$\eta-\eta'$ mixing angle \cite{Beneke:2003} & $\phi_\eta = 39.3\degree \pm 1.0\degree$
\\
CKM parameters \cite{CKMfitter}
                    & $A=0.807\pm0.018$
                \quad $\lambda=0.2265\pm0.0008$
                \quad $|V_{ub}/V_{cb}| = 0.881^{+0.011}_{-0.010}$
                \\&
                \quad $\phi_3(\gamma)=(67.4^{+2.8}_{-4.5})^\circ$
                \quad $\sin2\phi_1(\sin2\beta)
                        = 0.688^{+0.025}_{-0.024}$
\\
$B$ meson lifetimes \cite{Yao:2006px} & $\tau_{\barB^0} = 1.530\ps$
                $\tau_{B^-}=1.638\ps$
\\
$B\to P$ form factors \cite{Beneke:2003} &
              $F_0^{B\to\pi}(0) = 0.28$
              \quad $F_0^{B\to K}(0) = 0.34$
\end{tabular}
\end{ruledtabular}
\end{table}
\par
We also include hadronic uncertainty parameters defined in
\cite{Beneke:2000,Beneke:2006hg} : $X_H^{M}$ , $X_A^{M}$ ($M =
K,\etaP,\Kstar$), and $X_L^{\Kstar}$.
For simplicity, in the fit, we assume that $X_{A,H}^{K} =
X_{A,H}^{\etaP} \equiv X_{A,H}^{(K\etaP)}$ in $B\to K\etaP$ decays, and parameterize them as \cite{Beneke:2000}
\begin{eqnarray}
\begin{array}{l}
\displaystyle
X_{A,H}^{(K\etaP)} = \left[ 1+\rho_{A,H}\exp(i\phi_{A,H}) \right]
\log  \left(\frac{m_B}{\Lambda_h}\right),
\quad 0\le \rho_{A,H} \le 1.
\end{array}
\end{eqnarray}
In $B\to\phi\Kstar$ decays we have fixed $X_L = m_B/\Lambda_h$ and $X_H
= \log(m_B/\Lambda_h)$ (i.e., $\rho_H=\rho_L=0$),
because in these decays
the branching ratios are very insensitive to them.

As for NP effects, we use
$DB|\xi|$,
$DB|\xi'|$,
$\phi$,
$\phi',$
$\delta$,
$\delta'$
and $\theratio$
as the independent parameters.
We have constrained weak and strong phases to be
$|\phi^{(\prime)}| \le \pi$ and $|\delta^{(\prime)}| \le \pi/2$,
respectively.
In the fit, for simplicity we consider the two scenarios:
(i) NP-(A) for which $\xi'=0$ and
(ii) NP-(B) for which $\xi=0$.

\subsection{Experimental Data}
In the fit for $\barB\to \barK\etaP$ decays,
we use 7 observables including
3 averaged branching fractions,
3 direct CP violations
 and the $\barB^0 \to \barK^0\eta'$ indirect CP violation
$-\eta_{CP}S_{K\eta'}$.
Here $S_{K\etaP}$ is defined by
\begin{eqnarray}
S_{K\etaP} = \frac{2\Im(\lambda_f)}{1+|\lambda_f|^2},
\quad
\lambda_f = \frac{q}{p}\,
\frac{A(\bar{B}^0\to K^0_{S,L}\etaP)}{A(B^0\to K^0_{S,L}\etaP)},
\end{eqnarray}
with $q/p \simeq e^{-2i\phi_1}$ for $B_d^0$,
and $\eta_{CP}$ is the CP eigenvalue of $\ket{K^0_{S,L}\etaP}$.
The value of $-\eta_{CP}S_{K\etaP}$ should be close to
 $\sin2\phi_1$ in the SM.
The experimental data for $B\to K\etaP$ are listed
in Table~\ref{KetaP-values}.
\begin{table}[tbp]
\caption{World averages of observables for $B\to K\etaP$
are shown in the second column
\cite{Yao:2006px,Aubert:2005iy,Schumann:2006bg,Richichi:1999kj,Aubert:2005bq,Abe:2006xp,Aubert:2006fy,Chen:2000hv,Aubert:2006wv,Chen:2006nk,Barberio:2007cr}.
Upper limits are at 90\% CL.
In the third and fourth columns we have shown best fit values for
combined fit with corrections
received from $B\to\phi\Kstar$ annihilation.
Corresponding best fit parameters
are shown in Table \ref{combined-result} and in FIG.~\ref{DBfit}.
Best fit values of $B\to \phi\Kstar$
are shown in Table~\ref{phiK-values}.}\label{KetaP-values}
\begin{ruledtabular}
\begin{tabular}{cccc}
Observable & Experiment &
 \multicolumn{2}{c}{Combined Fit with $\phi\Kstar$ ann.}
\\
           &            &
  NP-(A) & NP-(B) \\
\hline
$\Br(B^+ \to \eta'K^+)\micro$ & $69.7^{+2.8}_{-2.7}$
                              & $68.9\pm2.2$   & $68.9\pm2.2$
\\
$\Br(B^0 \to \eta'K^0)\micro$ & $64.9\pm3.5$
                              & $66.3\pm2.1$  & $66.3\pm2.1$
\\
$\Br(B^+ \to \eta K^+)\micro$ & $2.2\pm0.3$
                              & $2.2\pm0.3$ & $2.2\pm0.3$
\\
$\Br(B^0 \to \eta K^0)\micro$ & ($<1.9$)
                              & $1.5\pm0.3$  & $1.5\pm0.3$
\\
$\ACP(B^+\to \eta'K^+)$ & $0.031\pm0.021$
                        & $0.031\pm0.021$   & $0.030\pm0.018$
\\
$\ACP(B^0\to \eta'K^0)$ & $0.09\pm0.06$
                        & $0.02\pm0.01$    & $0.02\pm0.01$
\\
$\ACP(B^+\to \eta K^+)$ & $-0.29\pm0.11$
                        & $-0.30\pm0.11$  & $-0.31\pm0.10$
\\
$\ACP(B^0\to \eta K^0)$ &  (N.A.)
                        & $-0.20\pm0.11$  & $-0.14\pm0.07$
\\
$\SCP{K^0\eta'}(B^0\to \eta' K^0)$ & $0.61\pm0.07$
                        & $0.70\pm0.03$ & $0.70\pm0.01$
\\
$\SCP{K^0\eta }(B^0\to \eta  K^0)$ & (N.A.)
                        & $0.89\pm0.17$ & $0.73\pm0.03$
\end{tabular}
\end{ruledtabular}
\end{table}
\par
For the $B\to \phi \Kstar$ decays we have 20 observables,
which include
$\barB^{0,-}\to\phi\barK^{*0,-}$ branching fractions (${\cal B}$),
polarization fractions ($f_L$, $f_\perp$),
and CP asymmetries
($A_{CP}^{\rm tot}$, $A_{CP}^0$, $A_{CP}^\perp$),
phases of polarized modes
($\phi_\para$, $\phi_\perp$),
and phase differences
($\Delta \phi_\para$, $\Delta\phi_\perp$),
where
$A_{CP}^{\rm tot}
 = (\sum_{\lambda}|\bar{A}_\lambda|^2-\sum_{\lambda}|A_{\lambda}|^2)/
   (\sum_{\lambda}|\bar{A}_\lambda|^2+\sum_{\lambda}|A_{\lambda}|^2)$,
$A_{CP}^\lambda
 = (|\bar{A}_\lambda|^2 - |A_{\lambda}|^2)
  /(|\bar{A}_\lambda|^2 + |A_{\lambda}|^2)$,\footnote{In BaBar measurements \cite{Aubert:2004xc}, instead of $A_{CP}^\lambda$, the asymmetries of $f_\lambda^{+1}$ and $f_\lambda^{-1}$ are defined. $f_\lambda^{+1}$ and $f_\lambda^{-1}$ are the polarization fractions measured in $\barB$ and $B$ decays, respectively.}
$\phi_\lambda = \arg(A_\lambda / A_0)$,
$\Delta\phi_\lambda = \frac{1}{2}\arg(\bar{A}_\lambda/\bar{A}_0\cdot
A_0/A_\lambda)$ with $\lambda =0, \para,\perp$.
%
%
The experimental data for the $B\to\phi\Kstar$ decays
are shown in Table~\ref{phiK-values}.
\begin{table}[tbp]
\caption{
 World averages and best fit values of observables for
 $\barB^0 \to \phi\barKstarz$ (upper)
 and $B^- \to \phi \Kstarm$ (lower)
 \cite{Aubert:2006uk,Bussey2006,Chen:2005zv,Yao:2006px,Barberio:2007cr,:2007br}.
In the third and fourth column, best fit values for combined
fit of $B\to K\etaP$ and $B \to \phi \Kstar$
for  NP-(A) and NP-(B) scenarios,
 including the contributions from $B\to \phi \Kstar$ annihilations,
 are shown.
Corresponding best fit parameters
are show in FIG.~\ref{DBfit} and Table~\ref{combined-result}.
Best fit values for $B\to K\etaP$
are shown in Table~\ref{KetaP-values}.}\label{phiK-values}
\begin{ruledtabular}
\begin{tabular}{cccc}
Observable & Experiment
 & \multicolumn{2}{c}{Combined Fit with $\phi\Kstar$ ann.}
\\
           &
& NP-(A) & NP-(B)
\\
\hline
$\Br_{\rm tot}{\micro}$ & $9.5\pm0.8$
                                        & $9.4 \pm0.6$ & $9.3\pm0.6$
\\
                      & $10.0\pm1.1$
                                        & $10.1\pm0.8$ & $9.4\pm0.6$
\\
$f_L$                 & $0.491\pm0.032$
                                        & $0.491\pm0.025$ & $0.493\pm0.024$
\\
                      & $0.50\pm0.05$
                                        & $0.499\pm0.028$ & $0.486\pm0.025$
\\
$f_\perp$             & $0.252\pm0.031$
                                        & $0.247\pm0.013$ & $0.239\pm0.021$
\\
                      & $0.20\pm0.05$
                                        & $0.243\pm0.010$ & $0.241\pm0.022$
\\
$\ACP^{\rm tot}$      & $-0.01\pm0.06$
                                        & $-0.03\pm0.05$   & $0.00\pm0.01$
\\
                      & $-0.01\pm0.08$
                                        & $0.02\pm0.03$   & $0.00\pm0.00$
\\
$\ACP^0$              & $0.02\pm0.07$
                                        & $-0.03\pm0.03$   & $0.01\pm0.01$
\\
                      & $0.17\pm0.11$
                                        & $0.05\pm0.07$   & $0.00\pm0.00$
\\
$\ACP^\perp$          & $-0.11\pm0.12$
                                        & $0.02\pm0.03$   & $0.00\pm0.00$
\\
                      & $0.22\pm0.25$
                                        & $0.05\pm0.07$   & $0.00\pm0.00$
\\
$\phi_{\parallel}$    & $2.37^{+0.14}_{-0.13}$
                                        & $2.34\pm0.09$ & $2.53\pm0.02$
\\
                      & $2.34\pm0.17$
                                        & $2.33\pm0.09$ & $2.52\pm0.02$
\\
$\phi_{\perp}$        & $2.36\pm0.14$
                                        & $2.49\pm0.06$ & $2.56\pm0.03$
\\
                      & $2.58\pm0.17$
                                        & $2.48\pm0.06$ & $2.55\pm0.03$
\\
$\Delta\phi_\parallel$& $0.10\pm0.14$
                                        & $0.02\pm0.06$ & $0.00\pm0.00$
\\
                      & $0.07\pm0.21$
                                        & $0.03\pm0.08$ & $-0.01\pm0.00$
\\
$\Delta\phi_\perp$    & $0.04\pm0.14$
                                        & $0.03\pm0.07$ & $0.00\pm0.00$
\\
                      & $0.19\pm0.21$
                                        & $0.04\pm0.09$ & $-0.01\pm0.00$
\end{tabular}
\end{ruledtabular}
\end{table}

\subsection{Combined Fits}
If we ignore the annihilation effects in $B\to \phi\Kstar$ decays,
the resulting $\chimin \gtrsim 170$ is too huge;
 i.e. we cannot have a reliable fitting result.
This is the fact that if the $B \to \phi\Kstar$ polarization anomaly
was mainly due to the tensor operators induced by
the Fierz transformation, then the NP effects would lead to too large
$B \to K\etaP$ branching ratios as compared with the data.
\par
Once the $B \to \phi \Kstar$ annihilation effects are included,
we can see that the $\chimin$ is drastically small.
In Table~\ref{combined-result}, we have summarized the best fit values
of the $\chimin$ and parameters.\footnote{
The errors of parameters in Table~\ref{combined-result} are obtained from the error matrix (covariance matrix) at the global minimum of $\chi^2$.
The error matrix is the inverse matrix of the curvature matrix of chi-square function with respect to its free parameters.
 The errors of best-fit values in Tables~\ref{KetaP-values} and \ref{phiK-values} are estimated from the same error matrices for each NP scenario.}
\begin{table}[tbp]
\caption{%
Best fit parameters obtained in
NP-(A) and NP-(B) with considering $B\to\phi\Kstar$ annihilations.
Numbers with $(*)$ indicates that they reach the upper or lower bound
in the parameter space.
Best fit values of $B\to K\etaP$ and $B\to\phi\Kstar$ are shown in
Tables~\ref{KetaP-values} and \ref{phiK-values}, respectively.
Weak phased NP parameters ($DB\xi$, $DB\xi'$) are plotted in FIG.~\ref{DBfit}.
}\label{combined-result}
\begin{ruledtabular}
\begin{tabular}{ccc}
\multicolumn{3}{c}{Combined fit with $\phi\Kstar$ annihilation effects}
\\
&  Scenario (A) & Scenario (B)
\\
 $\chimin/\dof$ & $9.8/17$
                & $15.5/17$
\\
\hline
$\theratio$    & $0.55\pm0.76$ (*)
               & $0.43\pm0.19$ (*)
\\
$\delta$ , $\delta'$
 & $\delta = +30\pm 85\degree$
 & $\delta'= -38\pm 70\degree$
\\
$\rho_A[K\etaP]$
 &  $1.00 \pm 0.33$ (*)
 &  $0.51 \pm 0.41$ (*)
\\
$\phi_A[K\etaP]$
 &  $117\pm17\degree$
 &  $53\pm47\degree$
\\
$\rho_H[K\etaP]$
 &  $0.11\pm0.67$ (*)
 &  $1.00\pm0.59$ (*)
\\
$\phi_H[K\etaP]$
 & $  69\pm156\degree$
 & $-167\pm25\degree$
\\
$\rho_A[\phi\Kstar]$
 & $0.57\pm0.04$
 & $0.55\pm0.02$
\\
$\phi_A[\phi\Kstar]$
 & $-96\pm3\degree$
 & $-85\pm3\degree$
\end{tabular}
\end{ruledtabular}
\end{table}
\begin{figure}[tbp]
\caption{%
Contour plots for
$\Delta \chi^2 \equiv \chi^2 - \chimin$
in $\Re (DB\xi)$ v.s. $\Im (DB\xi)$
[or $\Re (DB\xi')$ v.s. $\Im (DB\xi')$]
for the NP scenario-A [or NP scenario-B].
Allowed regions of
$\Delta\chi^2 < 1$,
$1 < \Delta\chi^2 < 4$ and
$4 < \Delta\chi^2 < 9$ are shown by dark, medium-dark and light-gray regions,
respectively.
``$\times$'' symbol indicates the location of the global minimum, $\chimin$.
The origin corresponds to the SM.
The circle at the origin indicates
the allowed upper-limit from the $B_s \to \mu^+\mu^-$ data.
}\label{DBfit}
\includegraphics{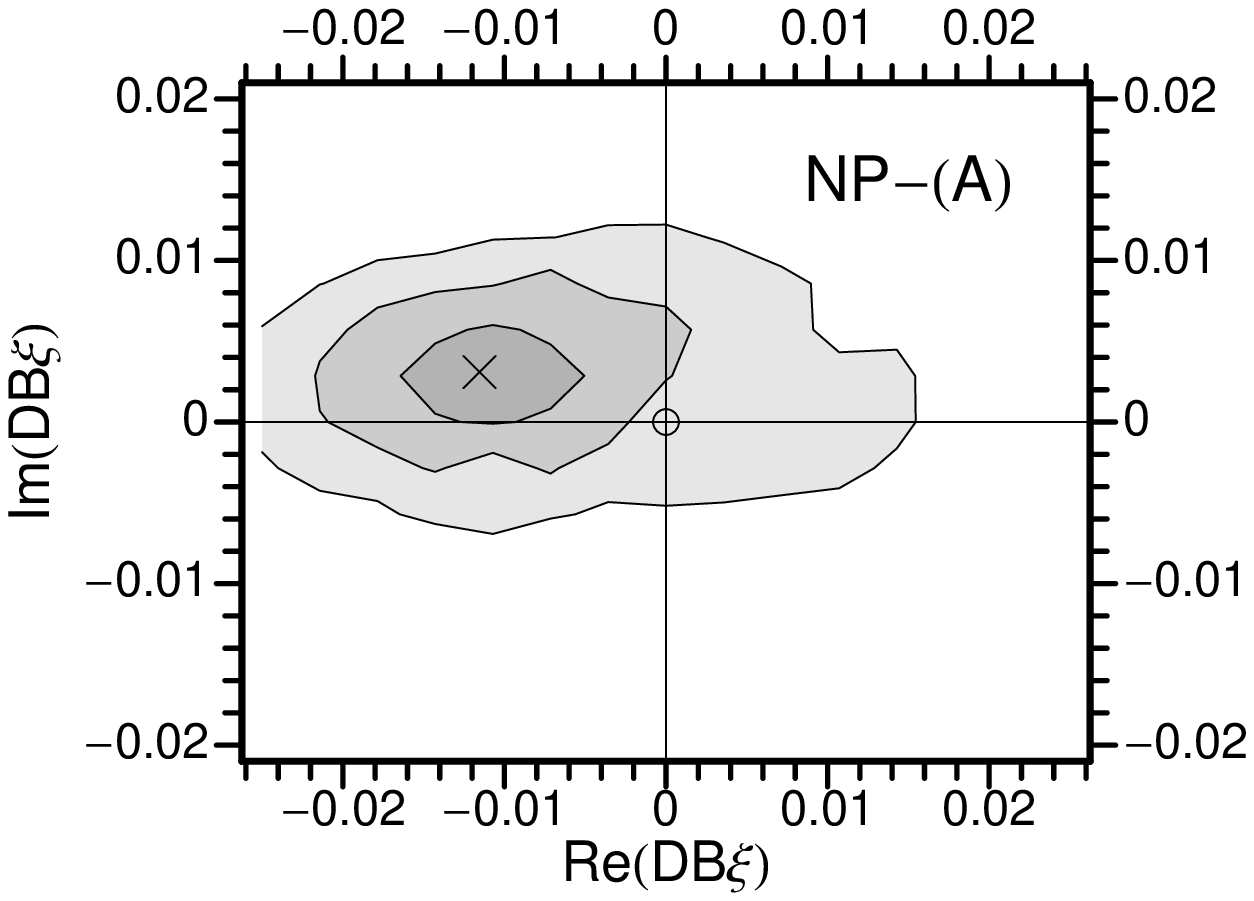}
\includegraphics{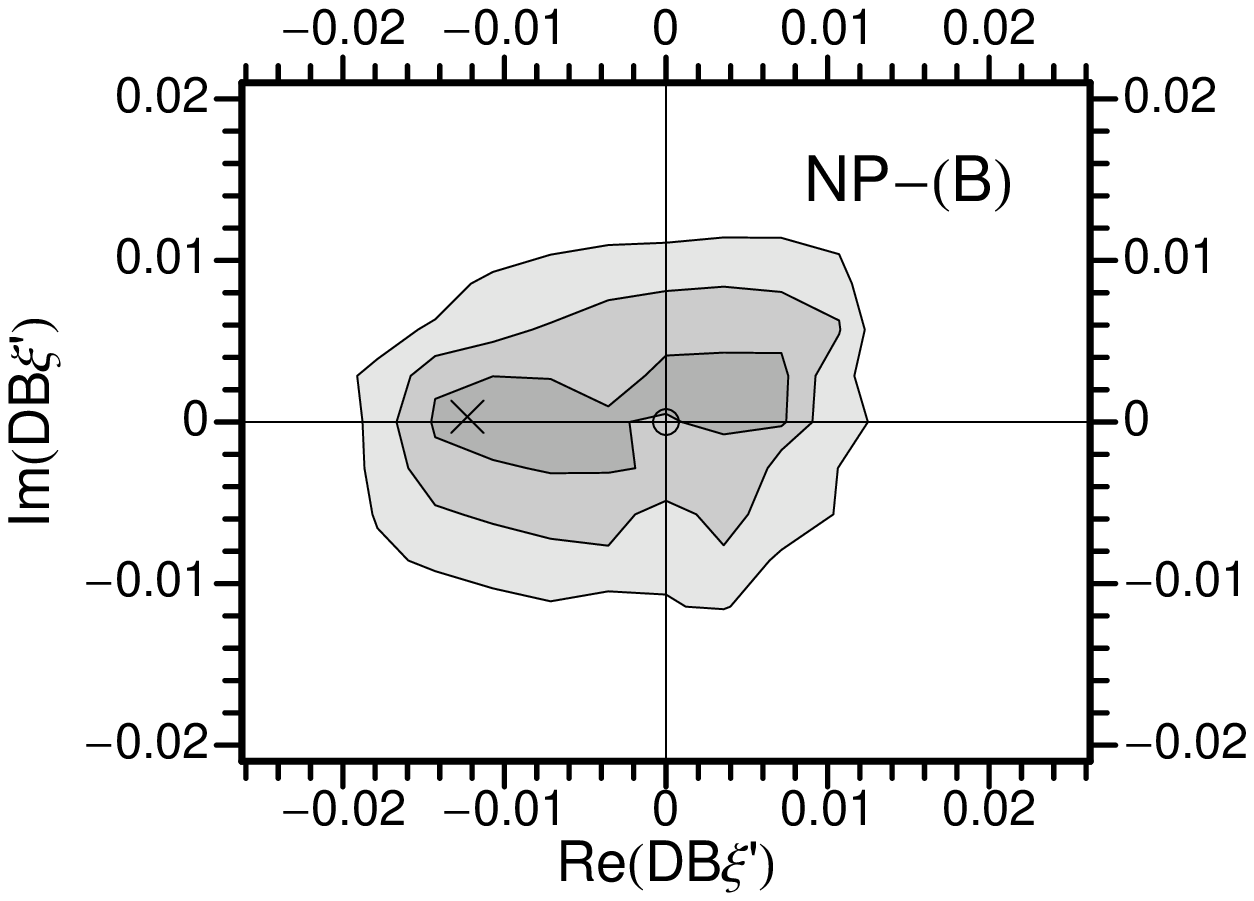}
\end{figure}
The resulting $\delta$ and $\delta'$ are consistent with zero,
which are also consistent with the fact that
the $\alpha_s$-corrections to the tensor operators are negligible.
The fitted results for $B\to K\etaP$ and $B\to \phi\Kstar$ are
collected in Tables~\ref{KetaP-values} and \ref{phiK-values}.
\par
The results obtained in \cite{Das-Yang},
where the weak annihilation effects are not included,
are given by\footnote{Here $\tilde{a}_{23}$ and $\tilde{a}_{25}$ are
defined by $\tilde{a}_{23} \equiv a_{23} + \frac{1}{2} a_{24}$ and
$\tilde{a}_{25} \equiv a_{25} +\frac{1}{2} a_{26}$,
 respectively \cite{Das-Yang}.
We have factored out the CKM factor.}
\begin{eqnarray}
\begin{array}{l}
|\tilde{a}_{23,{\rm DY}}| = 4.36^{+0.28}_{-0.18}\times 10^{-3},
\quad
|\tilde{a}_{25,{\rm DY}}| = 5.38^{+0.49}_{-0.31}\times 10^{-3},
\end{array}
\label{Das-Yang-result}
\end{eqnarray}
to be compared with our present upper bounds,
\begin{eqnarray}
|\tilde{a}_{23}| \le 7.1 \times 10^{-4} ,
\quad
|\tilde{a}_{25}| \le 6.1 \times 10^{-4},
\end{eqnarray}
which are extracted from FIG.~\ref{DBfit} and (\ref{phiKstar-MSSM}), and
are much smaller than the values in (\ref{Das-Yang-result}).
This indicates that the contributions of tensor operators induced from
the scalar/pseudoscalar operators in the MSSM-Higgs are too small to
explain the polarization puzzle, while the puzzle can be accommodated by
the weak annihilations effects.

\subsection{Consistency with SM and $B_s \to \mu^+ \mu^-$}
The current upper-bound for the branching fraction of $B_s \to
\mu^+\mu^-$ \cite{BsMuMu,Barberio:2007cr} at 90\% CL is
\begin{eqnarray}
{\cal B}(B_s \to \mu^+\mu^-) \le 7.5 \times 10^{-8}.
\label{BsMuMu-bound}
\end{eqnarray}
The branching fraction of $b \to s \lbar \l$ due to operators $O^{(\l)}_{i}$
(with $i=7,9,15,17,19,21$) for $\l=e,\mu,\tau$ \footnote{We have defined
$O^{(\l)}_i$ as $O_{7(9)}^{(\prime)} = \sbar (1-\gamma_5)b \lbar
(1\pm\gamma_5)\l$, and $O^{(\l)}_i$ ($i=15,\dots,21$) with replacements:
$\sbar(1\pm\gamma_5)s \to \lbar(1\pm\gamma_5)\l$ and $\sbar
\gamma^\mu(1+\gamma_5)s \to \lbar\gamma^\mu(1+\gamma_5)s$.} is given by
\begin{eqnarray}
\lefteqn{{\cal B}(B_s \to \mu^+\mu^-) }
\nonumber\\
&=&
\tau_{B_s}
\frac{G_F^2 m_{B_s}^3}{16\pi}
f_{B_s}^2
\left(\frac{m_{B_s}}{m_b+m_s}\right)^2
|V_{tb}^{}V_{ts}^*|^2 \sqrt{1-4\mhat^2}
\nonumber\\&&
\times
\left\{
(1-4\mhat^2)
\left|\cl_{15}-\cl_{17}+\cl_{19}-\cl_{21}\right|^2
+
\left|\cl_{15}+\cl_{17}-\cl_{19}-\cl_{21}
 + 2\mhat(\cl_{7}-\cl_{9})\right|^2
\right\},
\label{BsMuMu-formula}
\end{eqnarray}
where $\mhat \equiv m_\mu/m_{B_s}$,
$\cl_{i}$ are the Wilson coefficients of $O^{(\l)}_i$ at $\mu=m_b$,
and we have used
\begin{eqnarray}
\langle 0 | \sbar \gamma_5 b | \bar{B}_s \rangle
=
-i f_{B_s} \frac{m_{B_s}^2}{m_b + m_s}.
\end{eqnarray}
Here we note that
if RGE effects are not large,
$\cl_i \sim (m_\l/m_s) c_i$
and
\begin{eqnarray}
\begin{array}{lcl}
(c_{15} - c_{17} + c_{19} - c_{21})
&=& 2DA(\xi - \xi')
 = 2\left(1-\theratio\right)DB(\xi-\xi')
,
\\
(c_{15} + c_{17} - c_{19} - c_{21})
&=& -2DB(\xi + \xi') .
\end{array}
\end{eqnarray}
Using $B_s$'s lifetime $\tau_{B_s} = 1.437\ps$,
 mass $m_{B_s} = 5.366\GeV$,
decay constant $f_{B_s} = 215\pm25\MeV$,
quark masses $m_b=4.9\pm0.1\GeV$,
and $m_s = 145\pm25\MeV$,
we obtain upper bounds of $DB\xi^{(\prime)}$ as
\begin{eqnarray}
|DB\xi^{(\prime)}| \lesssim 9.2 \times 10^{-4}.
\end{eqnarray}
In FIG.~\ref{DBfit} we have shown the SM ($DB\xi=DB\xi'=0$)
and the upper bound of the NP effect constrained by the
$B_s \to\mu^+\mu^-$ decay as the origin and a small circle at the origin,
respectively.
In the figures we have also drawn the contours of $\Delta\chi^2 =1,4\text{ and }9$, where $\Delta\chi^2 \equiv \chi^2 - \chimin$.
The $B_s \to \mu^+\mu^-$ allowed region 
shares partly the $\Delta\chi^2 \le 1$ ($1\sigma$) region in NP-(B),
and is just outside of the $\Delta\chi^2 \le 4$ ($2\sigma$)
 region in the scenario NP-(A).
Since in both cases the $B_s\to\mu^+\mu^-$ data and the SM are located within contours where $\chi^2/\dof$ is sufficiently small,
we can safely conclude that our two scenarios are consistent with  
the data for $B_s \to \mu^+\mu^-$ decay and with the SM.
%
\par
In \cite{Faessler:2007br}, the authors discuss the scalar/pseudoscalar operators induced by R-parity violating interactions in the supersymmetric standard models.
Because they did not take into account the weak annihilation effects and possible constraints from $B\to K\etaP$, large contributions due to tensor operators to explain the $B\to\phi\Kstar$ polarization puzzle are required and therefore the estimated magnitudes of the effects of scalar/pseudoscalar operators are much larger than the upper bound of $B_s\to \mu^+\mu^-$.
%
\section{Summary}\label{sec-summary}
We have studied the scalar/pseudoscalar operators, and tensor
operators where the latter are obtained from scalar/pseudoscalar
operators by the Fierz transformation,
in $B\to\phi\Kstar$ and $B\to K\etaP$ decays.
We have considered the scalar/pseudoscalar operators induced by penguin
diagrams of MSSM neutral Higgs bosons.
%
\par
Without the weak annihilations in $B\to\phi\Kstar$,
we cannot obtain any reasonable solution
to explain both $B\to \phi\Kstar$ and $B \to
K\etaP$ decays simultaneously in the NP region
($-0.1 \le \theratio \le 1$) of the MSSM
induced by the neutral Higgs bosons.
%
%
Taking into account weak annihilation effects in $B\to \phi\Kstar$,
we obtain best fit results in good agreement
with the $B \to K\etaP$ and $B\to\phi\Kstar$ data.
From the fitted parameters we estimate
the magnitudes of the contributions due to NP tensor operators.
They are, however, much smaller than the results of \cite{Das-Yang},
which are introduced to explain
the $B\to\phi\Kstar$ polarization puzzle.
The polarization puzzle can be explained mostly
by weak annihilation effect,
as pointed out in \cite{Kagan:2004uw,Yang:2005tv,Beneke:2006hg}.
The contributions of NP operators are constrained mainly
by the fit of $B \to K \etaP$ data.
While our results may allow non-vanishing NP effects,
the data for the decays $B\to K\etaP$, $B\to\phi\Kstar$ 
are consistent with the $B_s\to \mu^+\mu^-$ data as well as
the SM prediction.
%
\par
Finally, we remark on the recently observed large longitudinal
polarization fraction $f_L$ in $B\to \phi K_2^*(1430)$
\cite{Aubert:2006uk}.
If tensor operators play an significant role in $B\to VT$
(where $T$ denotes a tensor meson) decays,
$f_L$ may significantly deviate from unity.
The current $B\to\phi K_2^*(1430)$ experiment seems to be consistent with
our conclusion
since in our analysis the effect due to tensor operators is
found to be very small.
However, in the present study we cannot exclude the possibility that sizable NP effects contribute directly to tensor operators, instead of scalar/pseudiscalar operators, and, moreover, a cancelation may take place between weak annihilations and contributions due to NP tensor operators in the $B\to\phi K_2^*(1430)$ decay.
For the point of view of the new physics, $B \to \phi K^*_2(1430)$
may be sensitive to the $B \to K_2^*$ tensor form factor
which can be further explored from the $B\to K^*_2(1430) \gamma$ decay.

\begin{acknowledgments}
We thank Andrei Gritsan for
many helpful comments on the manuscript. This work is partly supported by
National Science Council (NSC) of Republic of China
under Grants
NSC 96-2811-M-033-004 and
NSC 96-2112-M-033-MY3.
\end{acknowledgments}

\appendix
\section{Decay constants and Form Factors for $\eta$ and $\eta'$ mesons}\label{apdx-etap}
The $\ket{\eta}$ and $\ket{\eta'}$ meson states are defined as the
mixed states of $\ket{\eta_q}$ and $\ket{\eta_s}$, as stated in
(\ref{eta-mix}) \cite{Beneke:2002}.
In this section we summarize the notations in \cite{Beneke:2002}.
Decay constants $f_{\etaP}^{q,s}$ are given by
\begin{eqnarray}
\begin{array}{l}
f_\eta^q =  f_q \cos\phieta,
\quad
f_\eta^s = -f_s \sin\phieta,
\\
f_{\eta'}^q =  f_q \sin\phieta,
\quad
f_{\eta'}^s =  f_s \cos\phieta, \\
\end{array}
\end{eqnarray}
and in the same way, pseudoscalar densities $h_{\etaP}^{q,s}$ are defined as
\begin{eqnarray}
\begin{array}{l}
h_\eta^q =  h_q \cos\phieta,
\quad
h_\eta^s = -h_s \sin\phieta,
\\
h_{\eta'}^q = h_q \sin\phieta,
\quad
h_{\eta'}^s = h_s \cos\phieta, \\
\end{array}
\end{eqnarray}
where $h_{q,s}$ are defined by
\begin{eqnarray}
\begin{array}{l}
\displaystyle
h_q = f_q (m_{\eta}^2 \cos^2\phieta + m_{\eta'}^2 \sin^2\phieta)
 - \sqrt{2} f_s (m_{\eta'}^2 - m_{\eta}^2)\sin\phieta \cos\phieta,
\\
\displaystyle
h_s = f_s (m_{\eta'}^2 \cos^2\phieta + m_{\eta}^2 \sin^2\phieta)
 - \frac{1}{\sqrt{2}} f_q (m_{\eta'}^2 - m_{\eta}^2)\sin\phieta
 \cos\phieta.
\end{array}
\end{eqnarray}
$B\to \etaP$ form factors are defined as
\begin{eqnarray}
F^{B\to\etaP}
&=&
F_1 \frac{f_{\etaP}^q}{f_\pi} +
F_2 \frac{\sqrt{2}f_{\etaP}^q + f_{\etaP}^s}{\sqrt{3}f_{\pi}}
\label{BtoEta-formfactor}
\end{eqnarray}
and in the present paper we take $F_1 = F_0^{B\to\pi}(0)$ and $F_2 = 0$.
$f_\pi$ is the decay constant of the pion.
%
\section{Decay constants and Form factors in $B\to VV$ decays}\label{decay-form}
We have used
\begin{eqnarray}
\bra{\phi(q,\varepsilon_\phi)} V^\mu \ket{0}
 &=& f_\phi m_\phi \varepsilon_\phi^{\mu*},
\\
\bra{\barKstar(p_{\Kstar},\varepsilon_{\Kstar})} V_\mu \ket{\bar{B}(p_B)}
&=&
\frac{2}{m_B + m_{\Kstar}}
\epsilon_{\mu\nu\alpha\beta}
\varepsilon_{\Kstar}^{\nu*} p_B^\alpha p_{\Kstar}^\beta V(q^2),
\\
\bra{\barKstar(p_{\Kstar},\varepsilon_{\Kstar})} A_\mu \ket{\bar{B}(p_B)}
&=&i\left[(m_B + m_{\Kstar}) \varepsilon_{\Kstar\mu}^* A_1(q^2) -
     (\varepsilon_{\Kstar}^*\cdot p_B)(p_B + p_{\Kstar})_\mu
     \frac{A_2(q^2)}{m_B + m_{\Kstar}}\right]
\nonumber\\&&
-2im_{\Kstar} \frac{\varepsilon_{\Kstar}\cdot p_B}{q^2} q_\mu
\left[ A_3(q^2) - A_0(q^2)\right],
\label{form-pseudovector}
\end{eqnarray}
for current operators, and
\begin{eqnarray}
\bra{\phi(q,\varepsilon_1)}\sbar\sigma^{\mu\nu}s\ket{0}
&=& -i f_{\phi}^T
 (\varepsilon_1^{\mu*} q^\nu - \varepsilon_1^{\nu*} q^\mu),
\\
\bra{\barKstar(p',\varepsilon_2)} \sbar
 \sigma_{\mu\nu}(1+\gamma_5)b\ket{\barB(p)}
&=&
i\epsilon_{\mu\nu\rho\sigma}\varepsilon_2^{\nu*}p^\alpha p'^\beta
2T_1(q^2)
\nonumber\\
&&
+ \{\varepsilon_{2\mu}^*(m_B^2 - m_{\Kstar}^2) - (\varepsilon_2^* \cdot
p)(p+p')_\mu \} T_2(q^2)
\nonumber\\
&&
+
(\varepsilon_2^*\cdot p_B)
\left[q_\mu - \frac{q^2}{m_B^2 m_{\Kstar}^2}(p+p')_\mu\right] T_3(q^2),
\label{form-tensor}
\end{eqnarray}
for tensor operators.
In (\ref{form-pseudovector}) and (\ref{form-tensor}), $A_3(0) =
A_0(0)$, $T_1(0)=T_2(0)$ and
\begin{eqnarray}
A_3(q^2) = \frac{m_B + m_{\Kstar}}{2m_{\Kstar}} A_1(q^2) -
\frac{m_B - m_{\Kstar}}{2m_{\Kstar}} A_2(q^2).
\end{eqnarray}

\section{The coefficients $a_i^{p,h}$ corresponding to
 right-handed 4-quark operators}\label{app:ai_bi}

In (\ref{phiKstar-amp}), the expressions for effective parameters
$a_{11(12)}^{p,h}$ corresponding to
right-handed 4-quark operators are
\begin{eqnarray}\label{eq:app-ai}
 a_i^{p,h}(V_{1} V_{2})\! &=& \! \Bigg[ \bigg(c_i+{c_{i\pm1}\over
 N_c}\bigg)N_i(V_2) \nonumber\\
 && +{c_{i\pm1}\over N_c}\,{C_F\alpha_s\over
 4\pi}\Big(V_i^h(V_{2})+{4\pi^2\over
 N_c}H_i^h(V_{1} V_{2})\Big)+P_i^{p,h}(V_{2})
 \Bigg],
 {\hspace{0.5cm}}
 \end{eqnarray}
with $N_i(V_2) = 1$ for $i = 11,12$.
For $a^{p,h}_{13(14)}$, one should replace $c_i$ by $\bar{c_i}$,
and have $N_{13}(V_2)=1$, $N_{14}(V_2)=0$.
%
%
$V_i^h(V_2)$ account for vertex corrections, $H_i^h(V_1 V_2)$ for
hard spectator interactions with a hard gluon exchange between the
emitted meson and the spectator quark of the $B$ meson and
$P_i(V_2)$ for penguin contractions. The vertex corrections read
\begin{equation}\label{eq:vertex0}
   V_i^0(V_{2}) = \left\{\,\,
   \begin{array}{ll}
    {\displaystyle \int_0^1\!dx\,\Phi_{V_2}(x)\,
     \Big[ 12\ln\frac{m_b}{\mu} - 18 + g_T(x) \Big]} \,, & \qquad
     (i=\mbox{11,12}), \\[0.4cm]
   {\displaystyle \int_0^1\!dx\,\Phi_{V_2}(x)\,
     \Big[ - 12\ln\frac{m_b}{\mu} + 6 - g(1-x) \Big]} \,, & \qquad
     (i=13), \\[0.4cm]
   {\displaystyle \int_0^1\!dx\, \Phi_{v_2}(x)\,\Big[ -6 + h(x) \Big]}
    \,, & \qquad (i=14),
   \end{array}\right.
\end{equation}
and
\begin{equation}\label{eq:vertex+}
   V_i^+(V_{2}) = \left\{\,\,
   \begin{array}{ll}
    {\displaystyle \int_0^1\!dx\,\Phi_{b_2}(x)\,
     \Big[ 12\ln\frac{m_b}{\mu} - 18 + g_T(x) \Big]} \,, & \qquad
     (i=\mbox{11,12}), \\[0.4cm]
   {\displaystyle \int_0^1\!dx\,\Phi_{a_2}(x)\,
     \Big[ - 12\ln\frac{m_b}{\mu} + 6 - g_T(1-x) \Big]} \,, & \qquad
     (i=13) \\[0.4cm] 0
    \,, & \qquad (i=14),
   \end{array}\right.
\end{equation}
where $\Phi_V(x), \Phi_v(x), \Phi_a(x), \Phi_b(x)$, $g(x),\,h(x)$ and $g_T(x)$
are defined in \cite{Beneke:2003} and \cite{Beneke:2006hg}.

$H_i^h(V_1 V_2)$ have the expressions:
\begin{eqnarray}
H_{11}^0(V_1 V_2) &=& H_{12}^0(V_1 V_2) = {f_B f_{V_1} f_{V_2} \over
X_0^{(\overline{B} V_1, V_2)}}
  \int^1_0 d\rho {\Phi^B_1(\rho)\over \rho}\nonumber\\
   & & \times \int^1_0 d v \int^1_0 d u \,
 \Bigg( \frac{\Phi_{V_1}(v) \Phi_{V_2}(u)}{\bar u \bar v}
 + r_\chi^{V_1}
  \frac{\Phi_{v_{1}} (v) \Phi_{V_2}(u)}{u \bar v}\Bigg),
\label{eq:sepc01}
\\
  H_{13}^0(V_1 V_2) &=&- {f_B f_{V_1} f_{V_2} \over X_0^{(\overline{B} V_1, V_2)}}
  \int^1_0 d\rho {\Phi^B_1(\rho)\over \rho}\nonumber\\
   & & \times \int^1_0 d v \int^1_0 d u \,
 \Bigg( \frac{\Phi_{V_1}(v) \Phi_{V_2}(u)}{u \bar v}
 + r_\chi^{V_1}
  \frac{\Phi_{v_{1}} (v) \Phi_{V_2}(u)}{\bar u \bar v}\Bigg),
\label{eq:sepc02}
 \end{eqnarray}
$H_{14}^0(V_1 V_2)=0$ and
\begin{eqnarray}\label{eq:sepcp1}
  H_{11}^+(V_1 V_2) =H_{12}^+(V_1 V_2)
  &=& -{f_B f_{V_1}^\perp f_{V_2} \over X_+^{(\overline{B} V_1, V_2)}}
  \int^1_0 d\rho {\Phi^B_1(\rho)\over \rho} \int^1_0 d v \int^1_0 d u \,
  \frac{\Phi^\perp_{V_1}(v) \Phi_{b_2}(u)}{u \bar v^2},\nonumber\\
 H_{13}^+(V_1 V_2)
  &=& {f_B f_{V_1}^\perp f_{V_2} \over X_+^{(\overline{B} V_1, V_2)}}
  \int^1_0 d\rho {\Phi^B_1(\rho)\over \rho} \int^1_0 d v \int^1_0 d u \,
  \frac{\Phi^\perp_{V_1}(v) \Phi_{a_2}(u)}{\bar u \bar v^2},\nonumber\\
  H_{14}^+(V_1 V_2)
  &=& -{f_B f_{V_1} f_{V_2} \over 2X_+^{(\overline{B} V_1, V_2)}}
  \int^1_0 d\rho {\Phi^B_1(\rho)\over \rho} \int^1_0 d v \int^1_0 d u \,
  \frac{\Phi_{a_2}(v) \Phi^\perp_{V_2}(u) }{u \bar u\bar v},
 \end{eqnarray}
where
 \begin{eqnarray}
&& X^{({\overline B}^0 {V_1}, V_2)}_0 \nonumber\\
&&\ \ \ =
 \frac{f_{V_2}}{2m_{V_1}}\Bigg[(m_B^2-m_{V_1}^2-m_{V_2}^2)(m_B+m_{V_1})A^{B{V_1}}_1(m_{V_2}^2)
 -{4m_B^2p_c^2\over m_B+m_{V_1} }\,A^{B{V_1}}_2(m_{V_2}^2)\Bigg], \nonumber \\
&& X^{({\overline B}^0 {V_1}, V_2)}_+ = - f_{V_2} m_{V_2} \Bigg[
(m_B+m_{V_1})A^{B{V_1}}_1(m_{V_2}^2)\mp {2m_Bp_c\over m_B+m_{V_1}}\,
  V^{B{V_1}}(m_{V_2}^2) \Bigg],\end{eqnarray}
with $q = p_B - p_{V_1}\equiv p_{V_2}$. Here $\Phi^B_1(\rho)$ is one
of the two light-cone distribution amplitudes of the $\overline B$
meson~\cite{Beneke:1999}. $P_i^{h,p}$ are strong penguin contractions. We
obtain
\begin{eqnarray}
   P_{12}^{0,p}(V_2) &=& \frac{C_F\alpha_s}{4\pi N_c}\,
   \Bigg\{  (c_{12} +c_{14} )\!
    \sum_{i=u}^b \bigg[ {4 n_f \over 3}\ln \frac{m_b}{\mu}
    - (n_f -2) G_{V_2}(0) - G_{V_2}(s_c) - G_{V_2}(1) -\frac{8}{3}\bigg] \nonumber\\
    &&\!  +
    c_{11} \!
    \sum_{i=u}^b \bigg[ {8 \over 3}\ln \frac{m_b}{\mu}
    - G_{V_2}(0)  - G_{V_2}(1) + \frac{4}{3}\bigg]
    \Bigg\} ,
\label{eq:P-12}
\\
   P_{14}^{0,p}(V_2) &=& -\frac{C_F\alpha_s}{4\pi N_c}\,
   \Bigg\{  (c_{12} +c_{14} )\!
    \sum_{i=u}^b \bigg[ (n_f -2) \hat G_{V_2}(0) + \hat G_{V_2}(s_c) + \hat G_{V_2}(1)
    \bigg] \nonumber\\
    &&\!  +  c_{11} \!
    \sum_{i=u}^b \bigg[ \hat G_{V_2}(0) + \hat G_{V_2}(1)\bigg]
    \Bigg\} ,
\label{eq:P-14}
\end{eqnarray}
$P_{11}^{h,p}=P_{13}^{h,p}=P_{12}^{+,p}=P_{14}^{+,p}$, where
$s_i=m_i^2/m_b^2$ and the functions $G_{M_2}(s)$ and $\hat
G_{M_2}(s)$ are given by
\begin{eqnarray}
\begin{array}{l}
\displaystyle
 G_{M_2}(s) =   -4 \int^1_0 du\, \Phi_{V_2}(u)
 \bigg[ \int^1_0 dx\,x\bar x \ln (s-\bar u x\bar x-i\epsilon)\bigg]\,,
\\
\displaystyle
 \hat G_{V_2}(s) = -4 \int^1_0 du\, \Phi_{v_2}(u)
 \bigg[ \int^1_0 dx\,x\bar x \ln (s-\bar u x\bar x-i\epsilon)\bigg] \,.
\end{array}
\end{eqnarray}



\begin{thebibliography}{99}
%
\bibitem{Aubert:2006uk}
  B.~Aubert {\it et al.}  [BABAR Collaboration],
  Phys.\ Rev.\ Lett.\  {\bf 98}, 051801 (2007)
  [arXiv:hep-ex/0610073].

\bibitem{Chen:2005zv}
  K.~F.~Chen {\it et al.}  [BELLE Collaboration],
  Phys.\ Rev.\ Lett.\  {\bf 94}, 221804 (2005)
  [arXiv:hep-ex/0503013].

\bibitem{Bussey2006}
P. Bussey for the CDF Collaboration, ICHEP 2006.

\bibitem{:2007br}
  BABAR group,
  arXiv:0705.1798 [hep-ex]; 
  A.~V.~Gritsan,
{\it In the Proceedings of 5th Flavor Physics and CP Violation Conference (FPCP 2007), Bled, Slovenia, 12-16 May 2007, pp 001}
  [arXiv:0706.2030 [hep-ex]].





\bibitem{Cheng:2001aa}
  H.~Y.~Cheng and K.~C.~Yang,
  Phys.\ Lett.\  B {\bf 511}, 40 (2001)
  [arXiv:hep-ph/0104090].

%
%
\bibitem{Abe:2004mq}
  K.~Abe {\it et al.}  [BELLE-Collaboration],
  Phys.\ Rev.\ Lett.\  {\bf 95}, 141801 (2005)
  [arXiv:hep-ex/0408102].
\bibitem{Aubert:2006fs}
  B.~Aubert {\it et al.}  [BABAR Collaboration],
  Phys.\ Rev.\ Lett.\  {\bf 97}, 201801 (2006)
  [arXiv:hep-ex/0607057].

\bibitem{Li:2004ti}
  H.~n.~Li and S.~Mishima,
  Phys.\ Rev.\  D {\bf 71}, 054025 (2005)
  [arXiv:hep-ph/0411146].


\bibitem{Li:2003he}
  X.~Q.~Li, G.~r.~Lu and Y.~D.~Yang,
  Phys.\ Rev.\  D {\bf 68}, 114015 (2003)
  [Erratum-ibid.\  D {\bf 71}, 019902 (2005)]
  [arXiv:hep-ph/0309136].

\bibitem{Li:2004mp}
  H.~n.~Li,
  Phys.\ Lett.\  B {\bf 622}, 63 (2005)
  [arXiv:hep-ph/0411305].

\bibitem{Cheng:2004ru}
  H.~Y.~Cheng, C.~K.~Chua and A.~Soni,
  Phys.\ Rev.\  D {\bf 71}, 014030 (2005)
  [arXiv:hep-ph/0409317].
\bibitem{Ladisa:2004bp}
  M.~Ladisa, V.~Laporta, G.~Nardulli and P.~Santorelli,
  Phys.\ Rev.\  D {\bf 70}, 114025 (2004)
  [arXiv:hep-ph/0409286].

\bibitem{Kagan:2004uw}
  A.~L.~Kagan,
  Phys.\ Lett.\  B {\bf 601}, 151 (2004)
  [arXiv:hep-ph/0405134].

\bibitem{Yang:2005tv}
  K.~C.~Yang,
  Phys.\ Rev.\  D {\bf 72}, 034009 (2005)
  [Erratum-ibid.\  D {\bf 72}, 059901 (2005)]
  [arXiv:hep-ph/0506040].

\bibitem{Beneke:2006hg}
  M.~Beneke, J.~Rohrer and D.~Yang,
  Nucl.\ Phys.\  B {\bf 774}, 64 (2007)
  [arXiv:hep-ph/0612290].


\bibitem{Hou:2004vj}
  W.~S.~Hou and M.~Nagashima,
  arXiv:hep-ph/0408007.

\bibitem{Das-Yang}
  P.~K.~Das and K.~C.~Yang,
  Phys.\ Rev.\  D {\bf 71}, 094002 (2005)
  [arXiv:hep-ph/0412313].




\bibitem{Kagan:2004ia}
  A.~L.~Kagan,
  arXiv:hep-ph/0407076.

\bibitem{Alvarez:2004ci}
  E.~Alvarez, L.~N.~Epele, D.~G.~Dumm and A.~Szynkman,
  Phys.\ Rev.\  D {\bf 70}, 115014 (2004)
  [arXiv:hep-ph/0410096].



\bibitem{Chen:2006vs}
  C.~H.~Chen and H.~Hatanaka,
  Phys.\ Rev.\  D {\bf 73}, 075003 (2006)
  [arXiv:hep-ph/0602140].


\bibitem{Baek:2005jk}
  S.~Baek, A.~Datta, P.~Hamel, O.~F.~Hernandez and D.~London,
  Phys.\ Rev.\  D {\bf 72}, 094008 (2005)
  [arXiv:hep-ph/0508149].


\bibitem{Yang:2004pm}
  Y.~D.~Yang, R.~M.~Wang and G.~R.~Lu,
  Phys.\ Rev.\  D {\bf 72}, 015009 (2005)
  [arXiv:hep-ph/0411211].
\bibitem{Faessler:2007br}
  A.~Faessler, T.~Gutsche, J.~C.~Helo, S.~Kovalenko and V.~E.~Lyubovitskij,
  Phys.\ Rev.\  D {\bf 75}, 074029 (2007)
  [arXiv:hep-ph/0702020].

\bibitem{Chen:2005mka}
  C.~H.~Chen and C.~Q.~Geng,
  Phys.\ Rev.\  D {\bf 71}, 115004 (2005)
  [arXiv:hep-ph/0504145].

\bibitem{Chang:2006dh}
  Q.~Chang, X.~Q.~Li and Y.~D.~Yang,
  JHEP {\bf 0706}, 038 (2007)
  [arXiv:hep-ph/0610280].

\bibitem{Huang:2005qb}
  C.~S.~Huang, P.~Ko, X.~H.~Wu and Y.~D.~Yang,
  Phys.\ Rev.\  D {\bf 73}, 034026 (2006)
  [arXiv:hep-ph/0511129].

\bibitem{NHB}
  C.~S.~Huang and Q.~S.~Yan,
  Phys.\ Lett.\  B {\bf 442}, 209 (1998)
  [arXiv:hep-ph/9803366]; 
  C.~S.~Huang, W.~Liao and Q.~S.~Yan,
  Phys.\ Rev.\  D {\bf 59}, 011701 (1999)
  [arXiv:hep-ph/9803460]; 
  C.~S.~Huang, W.~Liao, Q.~S.~Yan and S.~H.~Zhu,
  Phys.\ Rev.\  D {\bf 63}, 114021 (2001)
  [Erratum-ibid.\  D {\bf 64}, 059902 (2001)]
  [arXiv:hep-ph/0006250]; 
  K.~S.~Babu and C.~F.~Kolda,
  Phys.\ Rev.\ Lett.\  {\bf 84}, 228 (2000)
  [arXiv:hep-ph/9909476]; 
  S.~R.~Choudhury and N.~Gaur,
  Phys.\ Lett.\  B {\bf 451}, 86 (1999)
  [arXiv:hep-ph/9810307].

\bibitem{Huang:2002ni}
  C.~S.~Huang and X.~H.~Wu,
  Nucl.\ Phys.\  B {\bf 657}, 304 (2003)
  [arXiv:hep-ph/0212220].
\bibitem{Cheng:2004jf}
  J.~F.~Cheng, C.~S.~Huang and X.~H.~Wu,
  Nucl.\ Phys.\  B {\bf 701}, 54 (2004)
  [arXiv:hep-ph/0404055].

\bibitem{Borzumati:1999qt}
  F.~Borzumati, C.~Greub, T.~Hurth and D.~Wyler,
  Phys.\ Rev.\  D {\bf 62}, 075005 (2000)
  [arXiv:hep-ph/9911245].


\bibitem{Beneke:2003}
  M.~Beneke and M.~Neubert,
  Nucl.\ Phys.\  B {\bf 675}, 333 (2003)
  [arXiv:hep-ph/0308039].
\bibitem{Beneke:2002}
  M.~Beneke and M.~Neubert,
  Nucl.\ Phys.\  B {\bf 651}, 225 (2003)
  [arXiv:hep-ph/0210085].


\bibitem{LCSR}
  A.~Ali and A.~S.~Safir,
  Eur.\ Phys.\ J.\  C {\bf 25}, 583 (2002)
  [arXiv:hep-ph/0205254];
  A.~Ali, P.~Ball, L.~T.~Handoko and G.~Hiller,
  Phys.\ Rev.\  D {\bf 61}, 074024 (2000)
  [arXiv:hep-ph/9910221].

\bibitem{Beneke:2000}
  M.~Beneke, G.~Buchalla, M.~Neubert and C.~T.~Sachrajda,
  Nucl.\ Phys.\  B {\bf 591}, 313 (2000)
  [arXiv:hep-ph/0006124].


\bibitem{CKMfitter} CKMfitter group
{\tt http://ckmfitter.in2p3.fr},
Results as of Summer 2007.

\bibitem{Yao:2006px}
  W.~M.~Yao {\it et al.}  [Particle Data Group],
  J.\ Phys.\ G {\bf 33} (2006) 1.
\bibitem{Barberio:2007cr}
  E.~Barberio {\it et al.}  [Heavy Flavor Averaging Group (HFAG)
                  Collaboration],
  arXiv:0704.3575 [hep-ex].


\bibitem{Aubert:2005iy}
  B.~Aubert {\it et al.}  [BaBar Collaboration],
  Phys.\ Rev.\ Lett.\  {\bf 94}, 191802 (2005)
  [arXiv:hep-ex/0502017].

\bibitem{Schumann:2006bg}
  J.~Schumann {\it et al.}  [Belle Collaboration],
  Phys.\ Rev.\ Lett.\  {\bf 97}, 061802 (2006)
  [arXiv:hep-ex/0603001].

\bibitem{Richichi:1999kj}
  S.~J.~Richichi {\it et al.}  [CLEO Collaboration],
  Phys.\ Rev.\ Lett.\  {\bf 85}, 520 (2000)
  [arXiv:hep-ex/9912059].

\bibitem{Aubert:2005bq}
  B.~Aubert {\it et al.}  [BABAR Collaboration],
  Phys.\ Rev.\ Lett.\  {\bf 95}, 131803 (2005)
  [arXiv:hep-ex/0503035].

\bibitem{Abe:2006xp}
  K.~Abe {\it et al.}  [Belle Collaboration],
  arXiv:hep-ex/0608033.

\bibitem{Aubert:2006fy}
  B.~Aubert {\it et al.}  [BaBar Collaboration],
  Phys.\ Rev.\  D {\bf 74}, 051106 (2006)
  [arXiv:hep-ex/0607063].

\bibitem{Chen:2000hv}
  S.~Chen {\it et al.}  [CLEO Collaboration],
  Phys.\ Rev.\ Lett.\  {\bf 85}, 525 (2000)
  [arXiv:hep-ex/0001009].

\bibitem{Aubert:2006wv}
  B.~Aubert {\it et al.}  [BABAR Collaboration],
  Phys.\ Rev.\ Lett.\  {\bf 98}, 031801 (2007)
  [arXiv:hep-ex/0609052].

\bibitem{Chen:2006nk}
  K.~F.~Chen {\it et al.}  [Belle Collaboration],
  Phys.\ Rev.\ Lett.\  {\bf 98}, 031802 (2007)
  [arXiv:hep-ex/0608039].



\bibitem{BsMuMu}
  D.~Tonelli  [CDF Collaboration],
{\it In the Proceedings of 4th Flavor Physics and CP Violation Conference (FPCP 2006), Vancouver, British Columbia, Canada, 9-12 Apr 2006, pp
001}
  [arXiv:hep-ex/0605038]; 
DO Collaboration, (V. Abazov etal), DO Note 5344-CONF (2007); 



\bibitem{Beneke:1999}
  M.~Beneke, G.~Buchalla, M.~Neubert and C.~T.~Sachrajda,
  Phys.\ Rev.\ Lett.\  {\bf 83}, 1914 (1999)
  [arXiv:hep-ph/9905312].


\bibitem{Aubert:2004xc}
  B.~Aubert {\it et al.}  [BABAR Collaboration],
  Phys.\ Rev.\ Lett.\  {\bf 93}, 231804 (2004)
  [arXiv:hep-ex/0408017].

\end{thebibliography}
\end{document}